\documentclass[iop]{emulateapj}
\usepackage{xcolor}

\usepackage{apjfonts}
\usepackage{graphicx}
\usepackage{epstopdf}

\usepackage{ulem}
\usepackage[utf8]{inputenc}
\usepackage[T1]{fontenc}
\usepackage{url}
\usepackage{mdwlist}
\usepackage{amsmath} % or simply amstext
\usepackage{enumitem}
\usepackage[colorlinks=true, linkcolor = blue, citecolor = blue]{hyperref}
\usepackage{lineno}
\makeatletter
\def\url@leostyle{%
 \@ifundefined{selectfont}{\def\UrlFont{\sf}}{\def\UrlFont{\small\ttfamily}}}
\makeatother
\begin{document}

\newcommand{\ls}{{_<\atop^{\sim}}}
\newcommand{\gs}{{_>\atop^{\sim}}}
\def \spose#1{\hbox  to 0pt{#1\hss}}  
\def \ls{\mathrel{\spose{\lower 3pt\hbox{$\sim$}}\raise  2.0pt\hbox{$<$}}}
\def \gs{\mathrel{\spose{\lower  3pt\hbox{$\sim$}}\raise 2.0pt\hbox{$>$}}}
\newcommand{\Ha}{\hbox{{\rm H}$\alpha$}}
\newcommand{\Hb}{\hbox{{\rm H}$\beta$}}
\newcommand{\Ovi}{\hbox{{\rm O}\kern 0.1em{\sc vi}}}
\newcommand{\OIII}{\hbox{[{\rm O}\kern 0.1em{\sc iii}]}}
\newcommand{\OII}{\hbox{[{\rm O}\kern 0.1em{\sc ii}]}}
\newcommand{\NII}{\hbox{[{\rm N}\kern 0.1em{\sc ii}]}}
\newcommand{\SII}{\hbox{[{\rm S}\kern 0.1em{\sc ii}]}}
\newcommand{\angstrom}{\textup{\AA}}
\newcommand\ionn[2]{#1$\;${\scshape{#2}}}

\font\btt=rm-lmtk10

%% ------------------------------------------------------------------
%% TITLE
%% ------------------------------------------------------------------

\title{Radio Morphology of Red Geysers}

\shorttitle{Radio Morphology of Red Geysers}%Looking for possible AGN signatures in Red geysers}
 
\shortauthors{Roy et al.}

%% ------------------------------------------------------------------
%% AUTHORS
%% ------------------------------------------------------------------

\author{Namrata Roy\altaffilmark{1}\dag, 
Emily Moravec\altaffilmark{2},
Kevin Bundy\altaffilmark{1,3}, 
Martin J. Hardcastle \altaffilmark{4},
G{\"u}lay G{\"u}rkan \altaffilmark{5},
Ranieri D. Baldi \altaffilmark{6,7},
Sarah K. Leslie \altaffilmark{8},
Karen Masters \altaffilmark{9},
Joseph Gelfand \altaffilmark{10},
Rogerio Riffel \altaffilmark{11,12}, 
Rogemar A. Riffel \altaffilmark{13,12},
Beatriz Mingo \altaffilmark{14}, 
Alexander Drabent \altaffilmark{15}}
%Kyle Westfall \altaffilmark{1,2},
%Rogerio Riffel\altaffilmark{6,7},
%Dmitry Bizyaev \altaffilmark{8},
%David V. Stark \altaffilmark{9},
%Rogemar A. Riffel\altaffilmark{10,7},

\altaffiltext{1} {Department of Astronomy and Astrophysics, University of California, 1156 High Street, Santa Cruz, CA 95064}\altaffiltext{\dag}{naroy@ucsc.edu}
\altaffiltext{2}{Astronomical Institute of the Czech Academy of Sciences, Bo\v cn\'i II 1401/1A, 14100 Praha 4, Czech Republic}
\altaffiltext{3}{UCO/Lick Observatory, Department of Astronomy and Astrophysics, University of California, 1156 High Street, Santa Cruz, CA 95064}
\altaffiltext{4}{School of Physics, Astronomy and Mathematics, University of Hertfordshire, College Lane, Hatfield, Hertfordshire AL10 9AB}
\altaffiltext{5}{Th{\"u}ringer State Observatory, Sternwarte 5, 07778 Tautenburg Germany}
\altaffiltext{6}{INAF - Istituto di Radioastronomia, Via P. Gobetti 101, I-40129 Bologna, Italy}
\altaffiltext{7}{School of Physics and Astronomy, University of Southampton, Southampton, SO17 1BJ, UK}
\altaffiltext{8}{Leiden Observatory, Leiden University, PO Box 9513, NL-2300 RA Leiden, The Netherlands}
\altaffiltext{9}{Haverford College,
Department of Physics and Astronomy,
370 Lancaster Ave, Haverford, PA 19041}
\altaffiltext{10}{New York University
Abu Dhabi, PO Box 129188, Abu Dhabi, UAE}
\altaffiltext{11}{Departamento de Astronomia, Instituto de F\'\i sica, Universidade Federal do Rio Grande do Sul, CP 15051, 91501-970, Porto Alegre, RS, Brazil}
\altaffiltext{12}{Laborat\'orio Interinstitucional de e-Astronomia - LIneA, Rua Gal. Jos\'e Cristin}
\altaffiltext{13}{Departamento de F\'isica, CCNE, Universidade Federal de Santa Maria, 97105-900, Santa Maria, RS, Brazil}
\altaffiltext{14}{School of Physical Sciences, The Open University, Walton Hall, Milton Keynes MK7 6AA, UK}
\altaffiltext{15}{Th{\"u}ringer Landessternwarte (TLS), Sternwarte 5, D-07778 Tautenburg, Germany}
%% ------------------------------------------------------------------
%% ABSTRACT 141/250 words
%% ------------------------------------------------------------------

\begin{abstract}

 We present 150 MHz, 1.4 GHz, and 3 GHz radio imaging (LoTSS, FIRST and VLASS) and spatially resolved ionized gas characteristics (SDSS IV-MaNGA) for 140 local ($z<0.1$) early-type ``red geyser'' galaxies. These galaxies have low star formation activity (SFR $\sim \rm 0.01\  M_{\odot} yr^{-1}$), but show unique extended patterns in spatially-resolved emission line maps that have been interpreted as large-scale ionized winds driven by active galactic nuclei (AGN). 
In this work we confirm that red geysers host low-luminosity radio sources ($\rm L_{1.4GHz} \sim 10^{22} W Hz^{-1}$). Out of 42 radio-detected red geysers, 32 are spatially resolved in LoTSS and FIRST, with radio sizes varying between $\sim 5-25$ kpc. Three sources have radio sizes exceeding 40 kpc. A majority display a compact radio morphology and are consistent with either low-power compact radio sources (``FR0'' galaxies) or ``radio-quiet quasars''.  They may be powered by small-scale AGN-driven jets which remain unresolved at the current $5''$ resolution of radio data. The extended radio sources, not belonging to the ``compact' morphological class, exhibit steeper spectra with a median spectral index of $-0.67$ indicating the dominance of lobed components. The red geysers hosting extended radio sources also have the lowest specific star formation rates, suggesting they either have a greater impact on the surrounding interstellar medium or are found in more massive halos on average. The degree of alignment of the ionized wind cone and the extended radio features are either 0$^{\circ}$ or 90$^{\circ}$, indicating possible interaction between the interstellar medium and the central radio AGN.

 %resolved ionized gas emission and kinematic maps with the radio morphology from LoTSS points to possible evidence of the relation between ionized wind and the central radio AGN.   
 
%%% mention steeper spctra in extended sources in \S 5.3. Mention the last line of abstract in \S 7. 

\end{abstract}

\keywords{Radio-quiet active galactic nuclei  -- galactic outflows -- radio jets}

\section{Introduction}

Active galactic nucleus (AGN) feedback has been proposed to be one of the most efficient ways to quench star formation and help maintain quiescence in massive galaxies and evoked to explain the enormous increase in the number of red galaxies since $z\sim2$. AGN feedback is often described as occurring in two different modes: ``quasar'' or ``radiative'' mode and ``maintenance'' or ``radio'' mode \citep{fabian12, morganti17, harrison18}. The ``quasar'' mode feedback, ushering in a rapid quenching phase during the early stage of a galaxy's lifetime, is associated with radiatively efficient luminous AGN or massive quasars. They release enormous amounts of energy to their surroundings via radiation from the accretion disk and drive powerful gas outflows that may remove gas altogether from the
galactic potential well \citep{cattaneo09, fabian12}. On the other hand, the ``radio'' mode feedback, predominant during the late stages of evolution, is thought to be powered by low to moderate luminosity AGN which are radiatively inefficient and accreting at a low rate. They deposit most of their energy to the surrounding medium via radio jets or winds, heating the gas and suppressing star formation \citep{binney95,ciotti01, croton06, bower06, ciotti07, ciotti10, mcnamara07, cattaneo09, fabian12, heckman14}. The radio mode feedback process has been directly observed  in galaxy groups and clusters \citep{mcnamara12}. Evidence for maintenance mode feedback in typical passive quenched galaxies (halo mass $\rm < 10^{13}~M_{\odot}$) have been rare. A few large-scale statistical studies of the local radio AGN population \citep{hardcastle19} and studies of individual galaxies showing radio AGN-driven outflows \citep{morganti05, nesvadba08} seem to suggest that the jet mechanical energy derived from the radio luminosity is enough to counterbalance the radiative loss of the hot gas and prevent cooling. %However, a systematic study directly investigating the role of radio AGN in a sample of early type galaxies (ETGs) which harbor multi phase gas outflows as well as exhibit very low star formation at present, has been lacking. 

%While the radio mode feedback process has been widely proposed in theory, direct observational evidence of this mode of feedback in typical passive galaxies has been limited, owing to the low-luminosity nature of the AGN. 

Low redshift integral field spectroscopy from the Sloan Digital Sky Survey-IV (SDSS-IV) Mapping Nearby Galaxies at Apache Point Observatory (MaNGA) survey \citep{bundy15} has recently revealed an interesting population of moderate mass (log M$_\star$/M$_{\odot}$ $\sim$ 10.5), red and quenched ($NUV -\ r > 5$) galaxies that may be useful in this regard. Known as ``red geysers'', these passive early-type galaxies possess unique optical emission and kinematic properties signalling galactic scale centrally-driven outflows \citep{cheung16, roy20}. The large scale winds of ionized gas, evident from the spatially resolved gas kinematics \citep{roy20}, aligns with a distinctive bi-symmetric enhancement in the spatial distribution of ionized gas, i.e. in H$\alpha$, [OIII] and [NII]. The observed ionized gas, traced by emission lines, is possibly ionized by post asymptotic giant branch (AGB) stars with some contribution from shocks, as evident from a combination of low ionization nuclear emission line regions (LINER) and Seyfert-like line ratios in spatially resolved BPT \citep[Baldwin, Phillips \& Terlevich,][]{baldwin} diagrams \citep{cheung16, roy20}. Using the Keck Echelette Spectrograph and Imager (ESI) instrument, we obtained high spectral resolution observations (R $\sim$ 8000 compared to 2000 in MaNGA) in two representative red geysers and found a systematic variation in the asymmetry of the emission line profiles. \cite{roy20} showed that the observed nature and the magnitude of asymmetry along with increased gas velocity dispersion are consistent with line-of-sight projections through a broad conical outflow. In addition, \cite{cheung16} performed detailed dynamical modeling of gas and stellar kinematics and concluded that the observed ionized gas velocities are too high to be in gravitationally bound orbits and can only be explained by an outflowing wind. These galaxies show very low star formation activity with average log SFR ($\rm M_{\odot}/yr) \sim -2$ using simultaneous SED fitting of GALEX+SDSS+WISE \citep{salim16} and present no visible signatures of dust lanes from ground based imaging. 

For a prototypical red geyser, \cite{cheung16} showed that the host galaxy has a radiatively-inefficient supermassive black hole which  was detected as a central radio point source. \cite{roy18} extended that analysis and used the Very large Array (VLA) Faint Images of the Radio Sky at Twenty-Centimeters \citep[FIRST, ][]{becker95} survey to measure stacked 1.4 GHz radio continuum flux from the entire red geyser sample. The study revealed that red geysers have significantly higher ($> \rm 5 \sigma$) radio continuum flux (in the stacked sample) and a three times higher radio-detection rate compared to the control samples. \cite{roy18} concluded that the red geysers host low-luminosity radio AGNs ($ \rm L_{\rm 1.4GHz} \sim 10^{22} - 10^{23} \ W/Hz$) which are energetically capable of driving sub-relativistic winds consistent with the MaNGA observations. Additionally \cite{roy21}, discovered a significant amount of cool gas (average $M_{\rm cool} \sim 10^{8}\ \rm M_{\odot}$) traced by sodium doublet absorption (NaD) in the red geyser sample, especially in those which are radio-detected according to FIRST. The spatial distribution of the cool gas lies spatially offset from the warm ionized gas component, as traced by H$\alpha$. The absorption line kinematics are observed to be redshifted on average ($\sim \rm 40 - 50 \ km \ s^{-1}$) in about 86\% of the radio red geysers, implying that the detected cool gas is inflowing into the galaxy and is possibly associated with fuelling the central radio AGN. The lack of any detectable star formation, the association with low luminosity radio-mode active galactic nuclei \citep{roy18}, signatures of large scale ($\sim 10\ \rm kpc$) ionized wind \citep{roy20} and their relatively high occurrence rate on the red sequence \citep[5-10 \%, ][]{cheung16}, make the red geysers a promising candidate for “maintenance'' or ``radio-mode'' feedback in typical quiescent galaxies.

While feedback from high luminosity radio-loud AGN, radio galaxies and radio Mpc-scaled jets have been extensively discussed and studied, feedback from low-luminosity radio AGN is less well understood. 
However recently, there have been a growing number of studies of the radio properties and morphology of ``radio-quiet'' sources and their relation with radio-mode AGNs hosting small scale jets that do not extend beyond the host galaxy. For example, \cite{jarvis19} presented 1-7 GHz high resolution radio imaging (VLA and e-MERLIN) for ten z$<$0.2 type-2 quasars which host ionized outflows based on broad [OIII] emission-line components. These ``radio-quiet quasars'' (RQQ) have low-to-moderate radio luminosities (log[$\rm L_{1.4GHz}/ W\  Hz^{-1}$] $\leq$ 24.5), exhibit extended radio structures in the scale of $\rm 1- 25$ kpc, and are consistent with being low power compact radio galaxies. The small-scale radio jets seem to be associated with ionized gas outflowing regions, indicating jet-interstellar medium (ISM) interaction on galactic scales \citep[similar to][]{venturi21}. 

\cite{capetti19}, on the other hand, explored the low-frequency (150 MHz) radio properties of similar compact low-luminosity (log[$\rm L_{150 MHz}/$ W Hz$^{-1}$] $\leq$ 22.5) radio AGN sources, known as Fanaroff-Riley class 0 (FR0), associated with nearby ($z< 0.05$) massive early-type galaxies. FR0 sources are typically unresolved with sizes $< \rm 3 - 6\ kpc$ and a few outliers showing a jetted morphology extending beyond 20 kpc. This class of sources represents the low end in size and radio power of small-scale AGN-jet population. Another set of ``galaxy-scale jets'' (GSJ) from 195 radio galaxies has been discovered by \cite{webster21} using LOFAR Two Metre Sky Survey \citep[LoTSS, ][]{shimwell19}. The radio emission from the GSJs extends to no larger than 80 kpc and are small enough to be directly influencing the evolution of the host galaxies. \cite{baldi18b, baldi21} studied high resolution ($< 0.2'')$ 1.5 GHz radio images for local active (LINERs and Seyferts) and inactive (HII and Absorption line galaxies) galaxies using e-MERLIN array. Investigating their radio morphology, they observed mostly radio cores with about one third of the detected sample featuring $\sim$ 1 kpc-scale radio jets. They concluded that the galaxies with LINER nuclei harbor radio sources which are scaled-down version of the FRI radio galaxies. Finally \cite{panessa19} has explored a wide range of possible mechanisms to understand the driver of the radio-quiet sources, starting from star formation, AGN driven
winds to free-free emission from photo-ionized gas and the innermost accretion disc coronal activity. 

Red geysers emerge as an interesting class of ETGs to study in the context of AGN-jet ISM interaction because they host low luminosity radio sources, exhibit suppressed star formation and show signatures of centrally driven outflows in ionized gas signatures. 
%in the context of these previously reported low luminosity radio-quiet sources featuring galactic-scale jet-ISM interaction. 
In this work we present the multi-frequency radio observations of 42 radio detected red geyser galaxies with 3 GHz using Very Large Array Sky Survey \citep[VLASS, ][]{myers18}, 1.4 GHz using FIRST and 150 MHz using LoTSS survey. Using a combination of radio and optical observations, we confirm that the observed radio emission is associated with radio mode AGN rather than star formation. We also investigate the spatial extent, morphology and spectral index of the radio emission from red geysers and explore the radio – ionized outflow connection.
In \S \ref{sample}, we report the red geyser sample and its unique identifying features. In \S\ref{surveys} we describe the radio and optical surveys used in this analyses. In \S \ref{result1}, \ref{result2} and \ref{sec:ionized} we present our results, which we then discuss in \S \ref{discussion}.
We summarise our conclusions in \S \ref{conclusion}. 

Throughout this paper, we assume a flat cosmological model with $H_{0} = 70$ km s$^{-1}$ Mpc$^{-1}$, $\Omega_{m} = 0.30$, and  $\Omega_{\Lambda} =0.70$, and all magnitudes are given in the AB magnitude system.

%--------------------------------------------------------------------
\section{Sample selection: Red Geysers}  \label{sample}

   \begin{figure*}
   \centering
   \includegraphics[width = \textwidth]{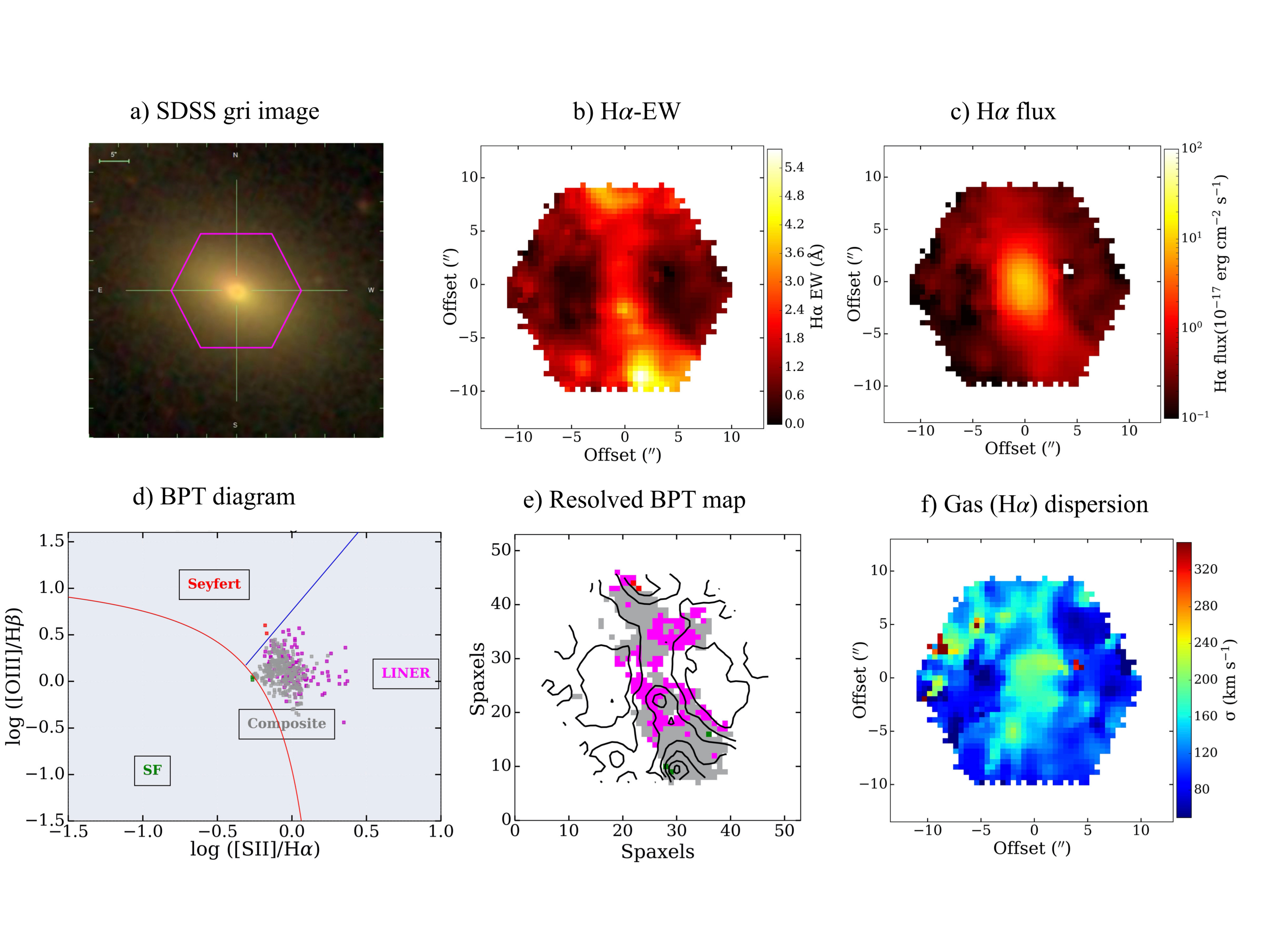}
   \caption{The spatially resolved emission and kinematic properties of an example red geyser (MaNGAID: 1-595166) as observed from SDSSIV-MaNGA. The upper left panel (a) shows the optical image of the galaxy from SDSS with the MaNGA IFU overlaid on top in magenta. In the other panels, we show the spatial distribution of H$\alpha$ equivalent width (b) and H$\alpha$ flux (c). The lower left panel (d) shows the spatially resolved [SII] BPT diagram showing spaxels with signal to noise $>$ 3. The spaxels are color-coded by the classification based on both the [NII] and [SII] BPT diagram. Almost all spaxels fall into the LINER/shock category (magenta color) while some spaxels are classified as ``composite'' from the [NII] BPT diagram (grey color). The lower middle panel (e) shows the spatial position of the BPT classified spaxels, with H$\alpha$ EW contours in black. The lower right panel (f) shows the gas velocity dispersion. The characteristic bi-symmetric pattern in H$\alpha$-EW map is particularly apparent. }%
   \label{fig:geysers_eg}
    \end{figure*}

This study of red geyser galaxies builds upon a series of papers in which we presented observations from radio (FIRST) and spatially resolved optical spectroscopic data (from SDSS-IV MaNGA survey and Keck ESI instrument) for a smaller sample of 84 red geysers. \cite{cheung16, roy18, roy20} and \cite{ roy21} have shown that the red geysers
are passive ETGs lying in the red sequence ($\rm NUV-r>5$) with ample amount of ionized and neutral gas present and they exhibit signatures of kpc-scale ionized winds driven out by a centrally located low-luminosity AGN. These galaxies show the widespread presence of ionized gas, traced by elevated flux of strong emission lines (e.g., H$\alpha$, [NII] and [OIII]) compared to other passive quenched galaxies, but with little ongoing star formation activity ($\sim 0.01$ $\rm M_\odot~yr^{-1}$). In this work we expand on \cite{roy18} by investigating the radio-detection of our updated sample of 140 red geysers and focus on the radio properties on three different radio bands $-$ 3 GHz, 1.4 GHz and 150 MHz. % sometimes surpassing $\rm 5 \times 10^{-16} \ erg \ cm^{-2} \ s^{-1}$. 
%They also show a unique bi-symmetric pattern in emission line equivalent width (EW) map, spatially coinciding with regions of enhanced flux, that has been interpreted as large scale AGN-driven wind. \cite{roy20}. 

The red geyser sample is visually selected from the SDSS IV-MaNGA survey (description of the survey in \S\ref{manga}) and has the following characteristic features \cite[see ][for details]{cheung16, roy18, roy20, roy21}:

\begin{itemize}
    \item Spheroidal galaxies (sersic index $>$ 3) with no visible disk component or dust lanes as observed by SDSS, red optical color (NUV $-$ r $>$ 5) and low star formation rate (average $\rm SFR \sim 10^{-2}~M_{\odot}~yr^{-1} $). Edge-on galaxies with axis ratio b/a $<$ 0.3 are discarded. 
    \item Bi-symmetric/bi-conical feature in spatially resolved EW map of H$\alpha$, [NII] and [OIII] emission lines. 
    \item Rough alignment (within $\rm \pm 10^{\circ}$) of the bi-symmetric feature with the ionized gas kinematic axis, but strong misalignment with stellar kinematic axis with the constraint that misalignment angle is not 90$^{\circ}$, 0$^{\circ}$ or 180$^{\circ}$. 
    \item High spatially resolved gas velocity values, typically reaching a maximum of $\pm \rm ~300~km~s^{-1}$, which are greater than the stellar velocity values by at least a factor of $\rm 4-5$.
    \item High gas velocity dispersion values, reaching about $\sim 220 - 250 \  \rm km~s^{-1}$ in distinct parts of the galaxy.
    \item Showing LINER or Seyfert type line ratios in the integrated BPT diagrams. 
\end{itemize}

An example red geyser is shown in Fig.~\ref{fig:geysers_eg}. The optical image (panel a) from SDSS shows spheroidal morphologies typical of these galaxies. The upper middle panel (b) shows the characteristic bi-symmetric feature in the $\rm H\alpha$ EW map. The upper right panel (c) shows the H$\alpha$ flux distribution which is extended in nature and shows enhanced values surpassing $\rm 5 \times 10^{-16} \ erg \ cm^{-2} \ s^{-1}$, a value quite high compared to typical passive galaxies. The regions of elevated H$\alpha$ flux coincides with the bi-symmetric EW pattern. This observed feature, which aligns with the gas velocity field, is believed to be tracing the putative wind cone. Panel d shows the spatially resolved [SII] BPT diagram for this red geyser in which only spaxels with signal to noise $>$ 3 are plotted. The lower middle panel (e) shows the spatial position of the spaxels in the spatially resolved BPT diagram, colored by their classification based on both the [NII] and [SII] BPT diagram \citep{kewley06}. Almost all spaxels fall into the LINER/shock category (magenta color) while some spaxels are classified as ``composite'' from the [NII] BPT diagram (grey color). The lower right panel (f) show the gas velocity dispersion map, traced by H$\alpha$, which is clumpy with values going up to $\sim 250 \ \rm  km\ s^{-1}$.

The sample of red geysers used in this work is derived from MaNGA Product Launch 9 (MPL-9) and consists of 140 galaxies, which account for $\approx6-8\%$ of the local quiescent galaxy population observed by MaNGA.

\section{Data Acquisition}  \label{surveys}

\subsection{MaNGA survey} \label{manga}

We use optical data primarily from the recently completed SDSS-IV MaNGA survey \citep{blanton17, bundy15, drory15, law15, yan16, sdss16}. 
MaNGA is an integral field spectroscopic survey that provides spatially resolved spectroscopy for nearby galaxies ($z\sim0.03$) with an effective spatial resolution of $2.5''$ (full width at half-maximum; FWHM). The MaNGA survey uses the SDSS 2.5 meter telescope in spectroscopic mode \citep{gunn06}
and the two dual-channel BOSS spectrographs \citep{smee13} 
that provide continuous wavelength coverage from the near-UV to the near-IR: $\rm3,600-10,000$ \AA. The spectral resolution varies from $\rm R\sim1400$ at 4000~\AA~ to $\rm R\sim2600$ at 9000~\AA. An $r$-band signal-to-noise $(S/N)$ of $\rm 4-8$~\AA$^{-1}$ is achieved in the outskirts (i.e., $\rm1-2~R_{e}$) of target galaxies with an integration time of approximately 3-hr. 
MaNGA has observed more than 10,000 galaxies with $\rm log~(M_\star/ M_{\odot})\geq 8$ across $\sim$ 2700 deg$^{2}$ over its 6~yr duration. In order to balance radial coverge versus spatial resolution, MaNGA observes two thirds of its galaxy sample to $\sim$ 1.5 R$_e$ and one third to 2.5 R$_e$. The MaNGA target selection is described in detail in \cite{wake17}.

The raw data are processed with the MaNGA Data Reduction Pipeline \citep[DRP,][]{law16}. 
In this work, we use the MaNGA sample and data products drawn from the MaNGA Product Launch-9 (MPL-9) and Data Release 16 \citep[DR16,][]{ahumada20}. 
We use spectral measurements and other analyses carried out by MaNGA Data Analysis Pipeline (DAP), specifically version 2.3.0. %\footnote{This version of the code will be made public in the upcoming SDSS Data Release 15 (DR15; Aguado et al. 2019, {\it submitted}). } 
The data we use in this work are based on DAP analysis of each spaxel in the MaNGA datacubes.  
The DAP first fits the stellar continuum of each spaxel to determine the stellar kinematics using the Penalised Pixel-fitting algorithm {\tt pPXF} \citep{cappellari04, cappellari17}  
and templates based on the MILES stellar library \citep{MILES}.  
The templates are a hierarchically clustered distillation of the full MILES stellar library into 49 templates.  This small set of templates provide statistically equivalent
fits to those that use the full library of 985 spectra in the MILES stellar library.  The emission-line regions are masked during this fit.
The DAP then subtracts the result of the stellar continuum modeling to provide a (nearly) continuum-free spectrum that is used to fit the nebular emission lines.  This version of the DAP treated each line independently, fitting each for its flux, Doppler shift, and width, assuming a Gaussian profile shape.  The final output from the DAP are gas and stellar kinematics, emission line properties and stellar absorption indices. All the spatially resolved 2D maps shown in the paper are outputs from the DAP with hybrid binning scheme. An overview of the DAP used for DR15 and its products is described by \cite{westfall19}, 
and assessments of its emission-line fitting approach is described by \cite{belfiore19}. All the integrated quantities reported in this paper are signal-to-noise weighted average taken over one effective radius. 
%This is different from the approach
%used by the DAP for DR15, which is to tie the velocities of all lines to a single value and to impose fixed flux ratios for the [OI], [OIII], and [NII] line doublets.  A detailed comparison of the results from the DR15 and MPL-5 versions of the DAP show that the different approach taken by the latter, and used for our analysis, has a negligible influence on our results.

We use ancillary data drawn from the NASA-Sloan Atlas\footnote{\url{http://www.nsatlas.org}} (NSA) catalog which reanalyzes images and derives morphological parameters for local galaxies observed in Sloan Digital Sky Survey imaging. It compiles spectroscopic redshifts, UV photometry \citep[from GALEX,][]{martin05}, stellar masses, and structural parameters. We have specifically used spectroscopic redshifts and stellar masses from the NSA catalog. The star formation rates are derived from \cite{salim16}, which utilizes GALEX-SDSS and WISE to perform UV-optical-IR spectral energy distribution (SED) fitting.

\subsection{LoTSS}  \label{lotss}

The LOFAR Two-metre Sky Survey \citep[LoTSS,][]{shimwell17, shimwell19} is an ongoing sensitive, high-resolution survey which will cover the whole northern sky with 3168 pointings in the frequency range between 120 and 168 MHz. The LoTSS first data release \citep[DR1,][]{shimwell19} covers 424 square degrees centred in the Hobby Eberly Telescope Dark Energy Experiment (HETDEX; Hill
et al. 2008) Spring Field region (right ascension 10h45m00s to
15h30m00s and declination $\rm 45^{\circ}00'00''$ to $\rm 57^{\circ}00'00''$) and contains over 300,000 sources with SNR $>$ 5. The median sensitivity is $\sim$ 71 $\mu$Jy/beam
and 95\% of the area in the DR1 release has an
rms noise level below 150 $\mu$Jy/beam.  The angular resolution
is 6$''$ and the positional accuracy is within 0.2$''$ for high signal-to-noise sources; the positional accuracy increases to $\sim$ 0.5$''$ for the faintest sources with
a flux density of less than 0.6 mJy. The source density is a factor of $\sim$10 higher than the most sensitive existing
very wide-area radio-continuum surveys such as
the NRAO VLA Sky Survey \citep[NVSS,][]{condon98},
Faint Images of the Radio Sky at Twenty-Centimeters \citep[FIRST,][]{becker95}, Sydney University Molonglo Sky Survey \citep[SUMSS,][]{bock99, mauch03}, and WEsterbork Northern Sky Survey \citep[WENSS,][]{rengelink97}.

The second LoTSS data release (DR2), to be released publicly in 2021, consists of two
contiguous fields at high Galactic latitude centered around
0h and 13h and covering approximately 5,700 square degrees (Shimwell et al. in preparation). DR2 provides
fully calibrated mosaics at the same $6''$ resolution as DR1, and images can also be obtained from individual LoTSS
pointings, outside the DR2 area.  
For the red geysers in the existing LoTSS coverage, including fields
not part of the DR2 release, we obtain the fluxes and sizes from
either (a) the internally released catalog (Shimwell et al. in prep)
or (b) similarly generated catalogues for small areas of individual
pointings around our target objects. In both cases, the flux scale
correction described by \cite{hardcastle21} and Shimwell et al. (in prep) is applied to the data so that
flux densities are as close as possible to the flux scale of
\cite{roger73}. The residual flux scale uncertainty lies between 5 and
10\% for the DR2 area \citep{hardcastle21} and is likely to be $\sim 10$\% for individual fields.

The total number of red geysers with currently available LoTSS data is 103 which is about 74\% of the parent 140 red geyser sample. The list of LOFAR detected red geysers is presented in Table~\ref{tab:summary}.

\subsection{FIRST survey}  \label{first}
The Very Large Array (VLA) Faint Images of the Radio Sky at Twenty Centimeters \citep[FIRST,][]{becker95} survey is a systematic survey over 10,000 square degrees of the North and South Galactic Caps at frequency channels centered at 1.36 GHz and 1.4 GHz. FIRST uses the VLA in B-configuration and achieves an angular resolution of 5$''$  and the survey is insensitive to structures larger than $\sim 60''$ as it is carried out in the VLA's B configuration. The source detection threshold is $\sim$ 1 mJy corresponding to a source density of $\sim$ 90 sources deg$^{-2}$. The astrometric accuracy of each source is 0.5 - 1$''$ at the source detection threshold. Since FIRST survey area was designed to overlap with the Sloan
Digital Sky Survey \citep[SDSS,][]{york00, abazajian09}, most MaNGA targets have FIRST data coverage. However, the 1 mJy threshold results in non-detections for most MaNGA galaxies. For each pointing center, there are twelve adjacent single field pointings that are co-added to produce the final FIRST image. Sources are extracted from co-added reduced images and fit by two dimensional Gaussians to derive peak flux, integrated flux densities, and size information \citep{becker95}. The current FIRST catalog is accessible from the FIRST search page. The full images are available from
\url{ftp://archive.stsci.edu/pub/vla\_first/data}. 

\subsection{VLASS}  \label{vlass}

The Very Large Array Sky Survey \citep[VLASS,][]{lacy16, myers18} is a community-driven initiative to carry out a synoptic radio sky survey using the Karl G. Jansky Very Large Array (VLA). VLASS will eventually use $\sim$ 5500 hours to cover the whole sky visible at the VLA ($\delta > -40$ deg) observing a total of 33,885 deg$^{2}$ at angular resolution of $\sim \rm 2.5''$. The data will be acquired in three epochs and will cover the frequency range 2–4 GHz in 2 MHz channels,
with calibrated polarimetry in Stokes I, Q and U, providing
wideband spectral and polarimetric data for a myriad of
targets and source types. The angular resolution is $\sim$ 2.5 arcsec and the survey is insensitive to structures larger than $\sim 30''$ as it is carried out in the VLA's B configuration. The 1 $\sigma$ sensitivity goal for a single pass is 120 $\mu$Jy while it is 69 $\mu$Jy when all three epochs are combined. Thus the VLASS is an all-sky radio sky survey with a unique combination
of high angular resolution, high sensitivity, full linear Stokes
polarimetry, time domain coverage, and wide bandwidth. Observing began in September 2017 and the survey will finish observing in 2024.

\section{Radio Detection and Characterization} \label{result1}

\subsection{Percentage of radio detection} \label{detection}

%\subsection{Host galaxy properties}
LoTSS DR2 footprint contains 93 of the parent sample of 140 red geyser galaxies. Additionally, ten more sources are contained within individual LoTSS pointings outside the DR2 area. Thus the total number of red geysers with currently available LOFAR data is 103, which is about 75\% of the complete red geyser sample. 34$~\pm~$6 out of those 103 sources ($\rm 33$\%$~\pm~$5.5\%) are found to be radio-detected at the frequencies observed by LOFAR ($\sim \rm 150\ MHz$), where quoted errors are obtained from standard Poisson statistics.

We cross-matched the FIRST catalog with our red geyser sample with a cross-matching radius of 10$''$. For visibly extended radio sources in the FIRST image, we extend the cross-matching radius up to a maximum of 1$'$, but restricting the central component of the detected source to lie within $< 10''$ of the galaxy center. We find that 29$~\pm~$5 out of 140 red geyser galaxies are detected at 1.4 GHz frequency with a detection fraction of $\rm 21$\%$~\pm~$3.5\%. This FIRST detection rate is roughly in agreement with our previous work \citep{roy18} which noted a 15\% radio detection rate from the FIRST survey. However \cite{roy18} used a preliminary red geyser sample of 84 sources drawn from an earlier MaNGA data release $-$ MPL-5. The current updated red geyser sample from MPL-9 has increased the total number of red geysers by 56 sources which results in an additional 17 FIRST detections for this work. %and the percentage of radio detection from the FIRST survey by $\sim \rm 6$\%.

Lastly, we cross-matched the parent red geyser sample with VLASS (3 GHz) survey using the same cross-matching radius as the FIRST survey and found 29 radio detections, which is a similar detection rate to FIRST. There are 13 red geysers which are detected in LOFAR but not in VLASS and FIRST. On the other hand, there are eight sources which are detected in FIRST and VLASS but are not within the LoTSS field of view. 
21$~\pm~$4 out of 140 red geysers ($\sim 15$\%$~\pm~$4\%) have simultaneous radio detections from all three surveys, which are used for calculation of spectral indices (discussed in \S \ref{spectral_index}). 42 $\pm$ 6 red geysers (30\% $\pm$ 4\%) are detected in at least one of LoTSS, FIRST or VLASS surveys, and these are listed in Table~\ref{tab:summary}.

    \begin{figure*} 
   \centering
   \includegraphics[width = \textwidth]{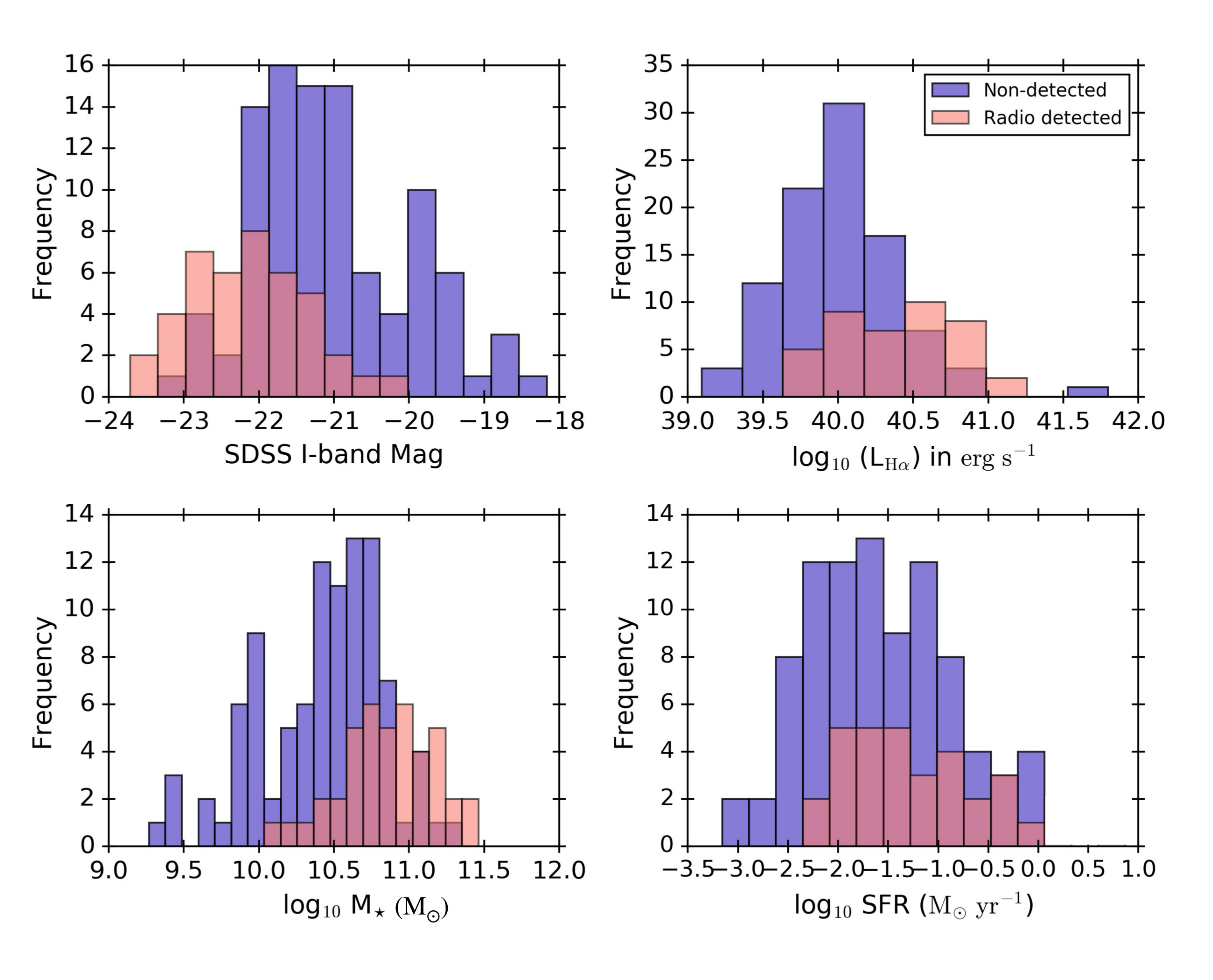}
   \caption{Comparison of global properties of radio-detected red geysers in salmon (detected in any of LOFAR, FIRST or VLASS survey) with non-radio detected red geysers (in blue). Histograms are shown for SDSS I-band magnitude, integrated H$\alpha$ luminosity, stellar mass and star formation rate. We see similar distributions of star formation rate but a substantial difference in the rest of the properties. }%
   \label{host}
    \end{figure*}

In order to understand how, if at all, the radio-detected red geyser galaxies intrinsically differ from those which are non-detected, we compare the host galaxy properties of the respective samples of interest. Fig.~\ref{host} shows the histograms of SDSS I-band magnitude (upper left), H$\alpha$ luminosity (upper right), stellar mass (lower left) and star formation rate (lower right) of non-radio detected red geysers (in blue) and 42 radio-detected red geysers (in salmon) $-$ detected in at least one of the LoTSS, FIRST or VLASS surveys. The SFR estimates are obtained from \cite{salim16} catalog, while the magnitude and the stellar mass reported here are acquired from the NSA catalog. We notice that the distributions of I-band magnitude, H$\alpha$ luminosity and stellar mass are quite different between the radio-detected and non detected sample. This is statistically confirmed by a Kolmogorov-Smirnov (KS) test which rejects the null hypothesis that the radio-detected and non-detected samples show similar distributions. This is shown by extremely small p values of $ \rm 1.48 \times 10^{-6}$, $1.84 \times 10^{-6}$, and $1.05 \times 10^{-6}$ for I-band magnitude, H$\alpha$ luminosity and stellar mass distributions respectively. However, for the SFR distributions, we cannot reject the null hypothesis at a level $< 1$\% ($p =0.06$), signifying similar distributions of radio and non-radio detected galaxies. The radio detected sources are in general brighter in the SDSS I-band (mean M$_{I} =  -22.5 $) and more massive (mean log$_{10}$ M$_{\star} \sim \rm 11\ M_{\odot}$) than the non-radio detected galaxies (mean M$_{I} =  -21$ and log$_{10}~\rm \ M_{\star} \sim \rm 10.5 \ M_{\odot}$). 
%This is in accordance with Sabater et al. (2018) which showed that the radio AGN activity shows a strong dependence on the stellar mass of the galaxy with the AGN fraction (L$\rm _{150MHz} \geq 10^{21} W\ Hz^{-1}$) reaching 100\% for most massive galaxies ($\rm \geq 10^{11} M_{\odot}$); thus the most massive galaxies are always switched on at some level. 
The H$\alpha$ luminosity in the radio-detected sample also tend to be higher on average (mean log$_{10} \rm \ L_{H\alpha} = \rm 40.5~erg~s^{-1}$) than the non-radio detected sample (log$_{10} \rm \ L_{H\alpha} = \rm 40 ~erg~s^{-1}$), possibly implying that the galaxies with higher amount of ionized gas and thus a more prominent H$\alpha$ bi-symmetric pattern are more likely to have enhanced radio emission. The SFR distribution, however, is unchanged irrespective of radio-detection indicating no underlying correlation between them.

\subsection{Radio loudness} \label{loudness}

\begin{figure}[h!!]
   \centering
   \includegraphics[width = 0.5\textwidth]{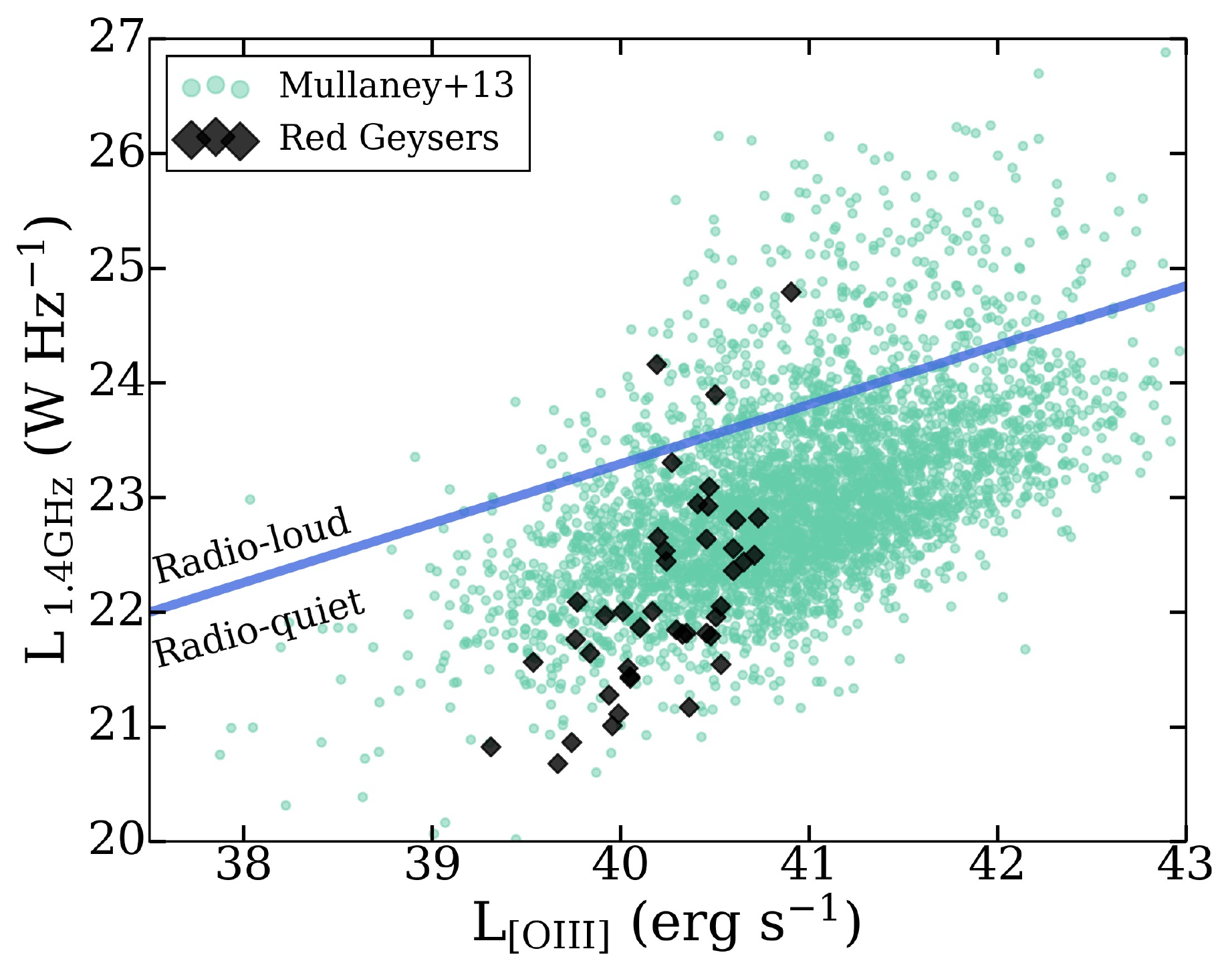}
   \caption{Radio luminosity measured at 1.4 GHz frequency vs. integrated [OIII] luminosity within central 3$''$ of red geyser galaxies (in black) and $z < 0.2$ AGN sample of \cite{mullaney13} (in green). The blue line divides the sample into `radio-loud' and `radio-quiet' according to \cite{xu99}. A large fraction of the red geysers are radio-quiet sources by this definition. }%
   \label{radio_quiet}
    \end{figure}

Fig.~\ref{radio_quiet} shows the radio luminosities at 1.4 GHz vs. the observed total [OIII] luminosities extracted from the central 3$''$ as observed by the SDSS fiber, of the 42 radio-detected red geysers (in black). Since LOFAR is most sensitive to fainter and extended radio emissions, it provides a more accurate estimate of the total flux density than FIRST. Hence we convert flux densities (S) measured at 144 MHz from LoTSS to 1.4 GHz for the LOFAR detected sources in order to compare to existing literature, assuming a spectral index ($\alpha$) of -0.7 using $\rm S_{\nu} \sim \nu^{\alpha}$ \citep{condon02}.  We use FIRST measured flux densities for the eight sources outside the LoTSS footprint. The radio luminosities thus obtained are compared to the \cite{mullaney13} z $<$ 0.2 AGN
population, plotted as green circles. The blue line marks the
division between ``radio-loud'' and ``radio-quiet'' sources from \cite{xu99}. A majority of our radio red geyser sample are classified as ``radio-quiet'' according to this definition with only three galaxies lying in the ``radio-loud'' regime. If we assume $\rm L_{1.4GHz} \sim L_{[OIII]}^{\beta}$ similar to \cite{xu99}, $\beta$ range from 0.53 to 0.60 in the red geyser sample with a mean value of 0.56. This is consistent with $\beta \sim \rm 0.5$ as reported in \cite{xu99}. 

In addition to using the criterion of \cite{xu99}, we compute the commonly used radio-loudness parameter R, the ratio of radio to optical brightness \citep{kellermann89}, to differentiate between ``radio-loud'' and ``radio-quiet'' sources in our sample. 
 Similar to \cite{ivezic02} and \cite{jarvis21}, we calculate R using the radio flux density at 1.4 GHz calculated in a similar way as mentioned above, and the SDSS i-band apparent magnitude using the following equation:

\begin{equation}
    R = 0.4(m_i - t)
\end{equation}

\noindent Here m$_{\rm i}$ is the Petrosian i-band apparent magnitude from SDSS DR16 \citep{ahumada20}. The Petrosian magnitudes used here recover essentially all of the flux of an exponential galaxy profile and about 80\% of the flux for a de Vaucouleurs profile. Here, t is the ``AB radio magnitude'' defined as 

\begin{equation}
    t = -2.5\ \rm  log \left(\frac{S_{\rm 1.4 GHz}}{3631 Jy} \right)
\end{equation}

\noindent where S$_{\rm 1.4 GHz}$ is the radio flux density (in Jy) measured at 1.4 GHz. We find that according to the $R$ parameter criterion, four out of 42 radio-detected red geysers would be classified as ``radio-loud'' with $R>\rm 1$. This includes the three galaxies which were identified as radio-loud according to \cite{xu99} and a fourth galaxy which was a borderline case $-$ lying just below the division line separating ``radio-loud'' and ``radio-quiet'' population (see Fig.~\ref{radio_quiet}). Although the $R$ parameter criterion is generally implemented in quasars with typical $R$ values for radio-loud sources going up to 2.8 \citep{ivezic02}, the $R$ values reported here simply quantifies the relative contribution of the radio luminosity over optical light.
93\% of the radio-detected red geysers are radio-quiet according to both criteria. 
This is consistent with \cite{roy18} which stated that the radio detected red geysers occupy the low-luminosity end ($\rm L_{1.4 GHz} < 10^{23}~W~Hz^{-1}$) of the radio population in the MaNGA quiescent galaxy sample.

The next section is dedicated to understanding the dominant mechanism responsible for the observed radio emission in these sources via physically motivated tests. %‘radio-quiet’ sources have radio emission primarily associated with the AGN, which we refer to as ‘radio AGN’, and not star formation. 

\subsection{Source of the radio flux: radio AGN or SF?} \label{radioagn}

    In Fig.~\ref{radio_quiet} we see that 39 out of 42 radio-detected red geysers in our sample would be classified as ``radio-quiet'' by the \cite{xu99} criterion. An important and significantly challenging follow-up question to address is whether the observed radio emission is associated with the central radio AGN or star formation. Although, we have predicted in \cite{roy18} that the radio continuum emission in red geysers is generally associated with central low-luminosity, radiatively inefficient radio-mode AGN, it is important to verify that interpretation in the light of other observations and in our increased sample of 42 radio detected red geysers. Star-forming galaxies (SFGs) emit at radio wavelengths primarily due to synchrotron emission from shocks associated with supernovae \citep{klein18}, %U. klein t al. 2018 - "Radio synchrotron spectra of star-forming galaxies'. 
    and hence their radio luminosity is expected to correlate broadly with the SFR. They generally display a diffuse clumpy radio emission not extending beyond the host galaxy with a steep spectral slope \citep{webster21, jarvis19}. On the other hand, radio emissions in the radio AGNs are primarily dominated by a jet originating from the central supermassive black
    hole. Unlike Mpc-scale radio jets in the centers of massive clusters and giant radio galaxies, the jets in low-luminosity AGN hosts are small scaled and confined near the very central region of the host galaxy \citep{jarvis19, venturi21, capetti20a, webster21} which often remains unresolved due to the low spatial resolution of various radio observations. This gives rise to compact or slightly extended radio sources with no visible lobes/ jets which are hard to distinguish from star forming galaxies. 
    
    In order to consider the possibility of star formation giving rise to the detected radio emission ($\sim \rm  10^{22}~W~Hz^{-1}$, Fig.~\ref{radio_quiet}), we need to detect a significant amount of star formation \citep[$\sim \rm 1-5~M_{\odot}$,][]{brown17} in the red geyser galaxies. If similar level of SF is not detected, we can rule out SF and attribute the observed radio emission to be from the central radio AGN. We consider three diagnostic plots to classify the radio-detected sources as either starforming or non-starforming galaxies:
    \begin{itemize}
        \item Identification based on WISE colors, particularly in W3–W2 \citep{yan13}. Star forming galaxies possess $\rm W3 - W2 > 0.3$. 
        \item Using emission line
diagnostics, in particular the ratio of [OIII] 5007 and H$\beta$ line fluxes, and that of [NII] 6584 and H$\alpha$ \citep{baldwin}, referred to as the ‘BPT’ method. Galaxies with $\rm log([OIII]/H\beta) < 0.5$ and $\rm log([NII]/H\alpha) < -0.3$ lying under the \cite{kauffman03} curve are star forming galaxies. 

        \item Using the relationship between
the 4000 \AA~break strength and radio luminosity per stellar mass \citep{best05}, hereafter referred to as the ‘D4000 vs $\rm L_{rad}/M$’ method. Galaxies with D4000$\leq$ 1.6 are dominated by young stellar populations and hence constitute star forming galaxies. 
    \end{itemize}

      \begin{figure*}[h!!]
   \centering
   \includegraphics[width = \textwidth]{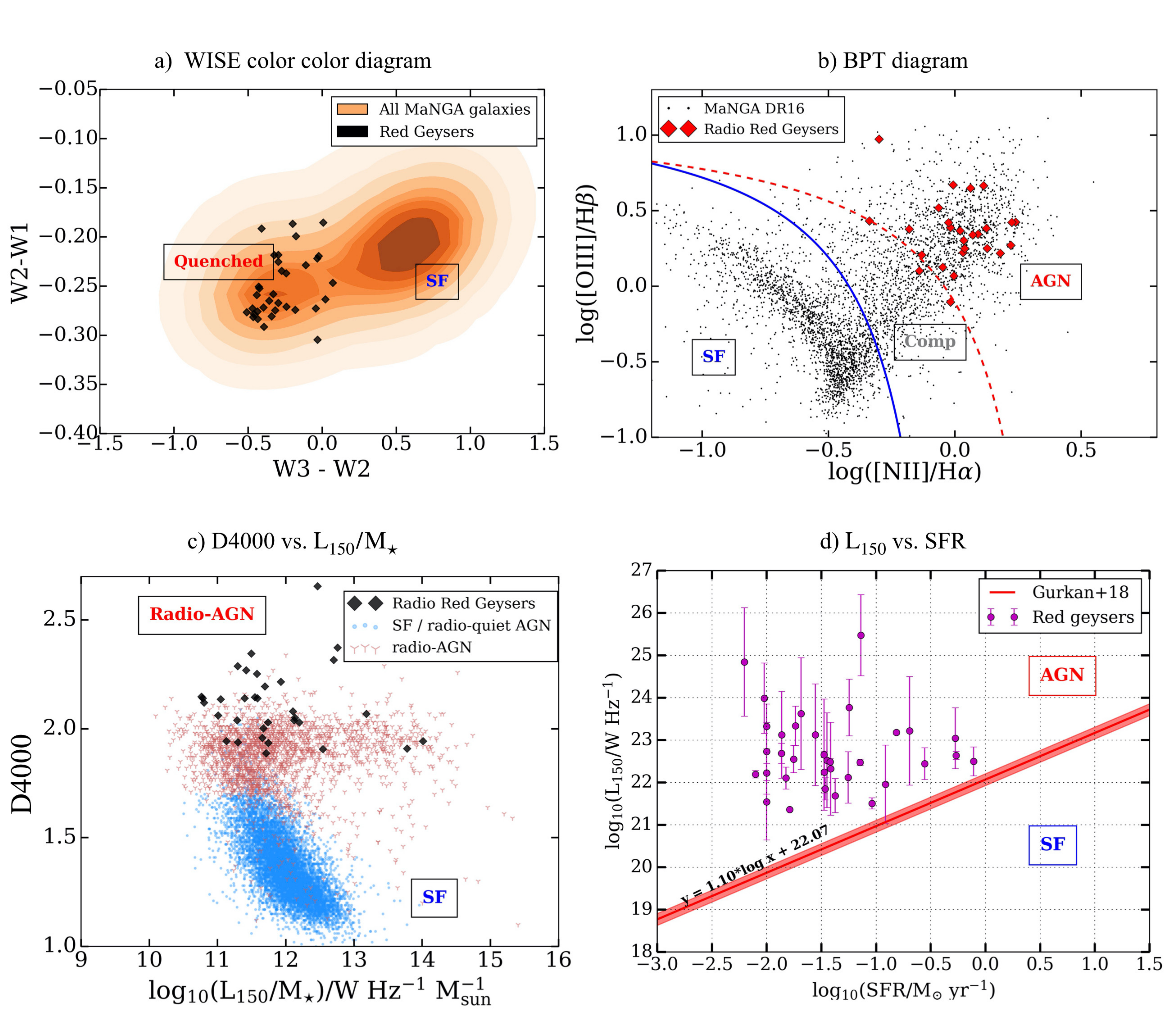}
   \caption{Location of the radio-detected red geyser sources on the different diagnostic plots used to separate the radio AGN from those galaxies where the radio emission is
        powered by SF. [Panel a] WISE W2-W1
        vs W3–W2 colour-colour diagnostic diagram. The radio detected red geysers are shown in black diamonds, while the orange contours are the wise-detected MaNGA galaxies. [Panel b] The [OIII]/H$\beta$ versus [NII]/H$\alpha$ emission line ratio diagnostic diagram from \cite{baldwin}, known as the 'BPT-diagram', for the radio-detected red geysers (in red) and MaNGA galaxies from data release 16 (in black). [Panel c] The ‘D4000 versus L$_{\rm 150 MHz/M_{\star}}$’ method, developed by \cite{best05}, for the radio-detected red geysers (in black). They are compared to the SF (blue) and AGN population (red) from the LoTSS (DR1) detected SDSS (DR7) galaxies by \cite{sabater19}. [Panel d] Radio luminosities vs. star formation rate for the LoTSS-detected red geysers (magenta) compared to the \cite{gurkan18} measured low frequency radio luminosity - star formation relation (red). All the red geysers lie above the said line, indicating that the radio emission observed in the red geysers are too high to be explained from the amount of star formation detected in these galaxies. }%
   \label{fig:quenched}
    \end{figure*}

Specific WISE mid-infrared colors can be used to separate galaxies with and without star formation. Thus, star-forming galaxies separate from the typical hosts of radio AGN in their WISE colours, particularly in W2–W3 \citep[4.6 to 12 micron color,][]{yan13}.
%Thus, for our sample of interest, this can confirm the absence of star formation and rule out SF being the dominant cause for the observed radio emission. 
Fig.~\ref{fig:quenched} upper left (panel a) shows a plot of W1–W2 versus W2–W3 mid-infrared WISE colors for 42 red geysers which are radio-detected in at least one of the LoTSS, VLASS or FIRST surveys (in black). The background orange contours represents the WISE colors for all galaxies in SDSS-MaNGA DR16 sample. The contour clearly indicates a bimodal distribution in the color space representing the star forming and quenched galaxy population. Star forming galaxies mostly occupy regions with $\rm 0.3<W3-W2<0.9$, while the quenched population has $\rm -0.6<W3-W2<0.0$. The radio-detected red geysers lie in the quenched part of the diagram which confirms the passive nature of these galaxies.

Since our red geyser targets possess strong emission lines, a common and useful method to separate SFG from AGN-hosts is through the ionization of the gas via Baldwin-Phillips-Terlevich diagram \citep[BPT,][]{baldwin}. By observing the relative strengths of four emission lines, namely [OIII]/H$\beta$ and [NII]/H$\alpha$, we can separate SFG and AGN-host galaxies based on the hardness of their ionizing spectrum, that in turn, drives the relative fluxes of different emission lines. This leads to 
the AGN-host galaxies to occupy a separate region in the diagram from the SF galaxies with the \cite{kewley06} and \cite{kauffman03} demarcation lines in between. 
Fig.~\ref{fig:quenched} upper right (panel b) shows the BPT diagram of all galaxies from MaNGA Data Release 16 in black. The red diamonds indicate the radio detected red geysers (detected in at least one radio band). The red geysers land in either the LINER or AGN regions of the BPT diagram and show no indication of SF activity. The absence of SF through the BPT diagram provides an useful diagnostic, as this confirms and re-iterates the quiescent ``red and dead'' nature of the galaxies and indicates that the possible source of the radio emission is an AGN.

The ‘D4000 vs L$_{\rm rad}$/M’ method for identifying radio AGN was developed by \cite{best05}. The parameter D4000 is the strength of the 4000 \AA~break in the
galaxy spectrum, and L$_{\rm rad}$/M$_{\star}$ is the ratio of radio luminosity (measured in a specific radio band) to stellar mass. This identification process is constructed on the basis that SFGs with a wide range of star formation histories occupy the same region in this plane since both L$_{\rm rad}$/M and D4000 depend broadly on the specific star formation rate of the galaxy. On the other hand, radio-loud AGN have enhanced values of $\rm L_{rad}$ and are thus separable on this plane. Among the low-luminosity radio sources, low D4000 value would distinguish galaxies with active star formation which would possibly be the dominant cause behind the observed radio emission in those sources. This identification method, later implemented with slight modifications by \cite{kauffman08, sabater19}, has been generally successful with few cases of mis-classification. 

In Fig.~\ref{fig:quenched} lower left panel (c), we plot D4000 vs. L$_{150}$/M$_{\star}$ for the radio-detected red geysers in black. We use only the 34 LOFAR-detected sources in this analyses utilizing the flux measurements from the 150 MHz band, which are then compared with existing sources from the literature.  We overplot the radio sources from SDSS DR7 from \cite{sabater19} in the background. The data points are color-coded in blue circles and red arrows which represents SFG and radio AGN respectively, classified using a combination of diagnostics \cite[see][]{sabater19}. In general, sources with average D4000 value exceeding $\sim $ 1.7 do not exhibit enough active star formation to show substantial radio emission due to supernovae/ stellar activity. Hence radio sources with L$\rm _{150}$/M$_{\star} > \rm 11~W~Hz^{-1}~M_{\odot}^{-1} $ and D4000 $> 1.7 $ are predominantly radio AGN. All the red geysers in our sample land in the radio AGN portion of the diagram, as they have a relatively old stellar population with D4000 exceeding 2.0.  

In addition to these three diagnostic plots, we also show the relation between radio luminosity (from LOFAR at $\sim$150 MHz) and SFR (panel d) for the LOFAR-detected red geysers sample (in magenta). As mentioned previously, star formation rate is expected to correlate with radio luminosity in star forming galaxies, due to synchrotron emission from supernovae shocks. \cite{gurkan18} studied low frequency radio luminosity - SFR relation on a large sample of SDSS galaxies and found the best fit (single power law) relation to be:
\begin{equation}
    \rm log_{10} (L_{150}) = 1.07 \pm 0.01 \times log_{10}(SFR) + 22.07 \pm 0.01
\end{equation}

\noindent The above relation is shown as a red line in Fig.~\ref{fig:quenched}. Any galaxy lying above this relation posses an excess amount of radio emission, too high to be  produced from the corresponding level of SF. We find that all the red geysers lie above the \cite{gurkan18} relation, which indicates that the radio emission is consistent with galaxies hosting radio AGNs.

\section{Radio Morphology}  \label{result2}

      \begin{figure*}
   \centering
   \includegraphics[width = \textwidth]{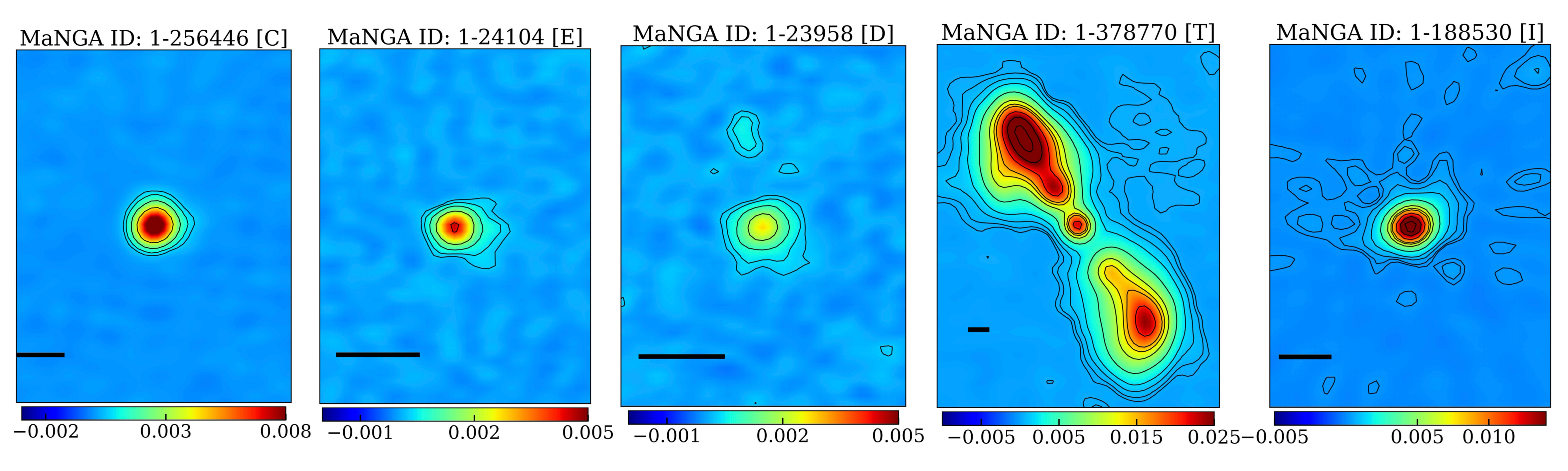}
   \caption{The LoTSS images of five example red geysers, classified in different radio morphology classes as mentioned in square brackets. Galaxy with MaNGAID: 1-256446 is classified as ``compact'' source, 1-24104 as ``extended'', 1-23958 as ``double'', 1-378770 as ``triple'' and 1-188530 as ``irregular'' source. The black horizontal scale bar in each panel indicates 20 kpc length scale. }%
   \label{example_lofar}
    \end{figure*}

% radio over optical : comparison between diff bands

      \begin{figure*}
   \centering
   \includegraphics[width = \textwidth]{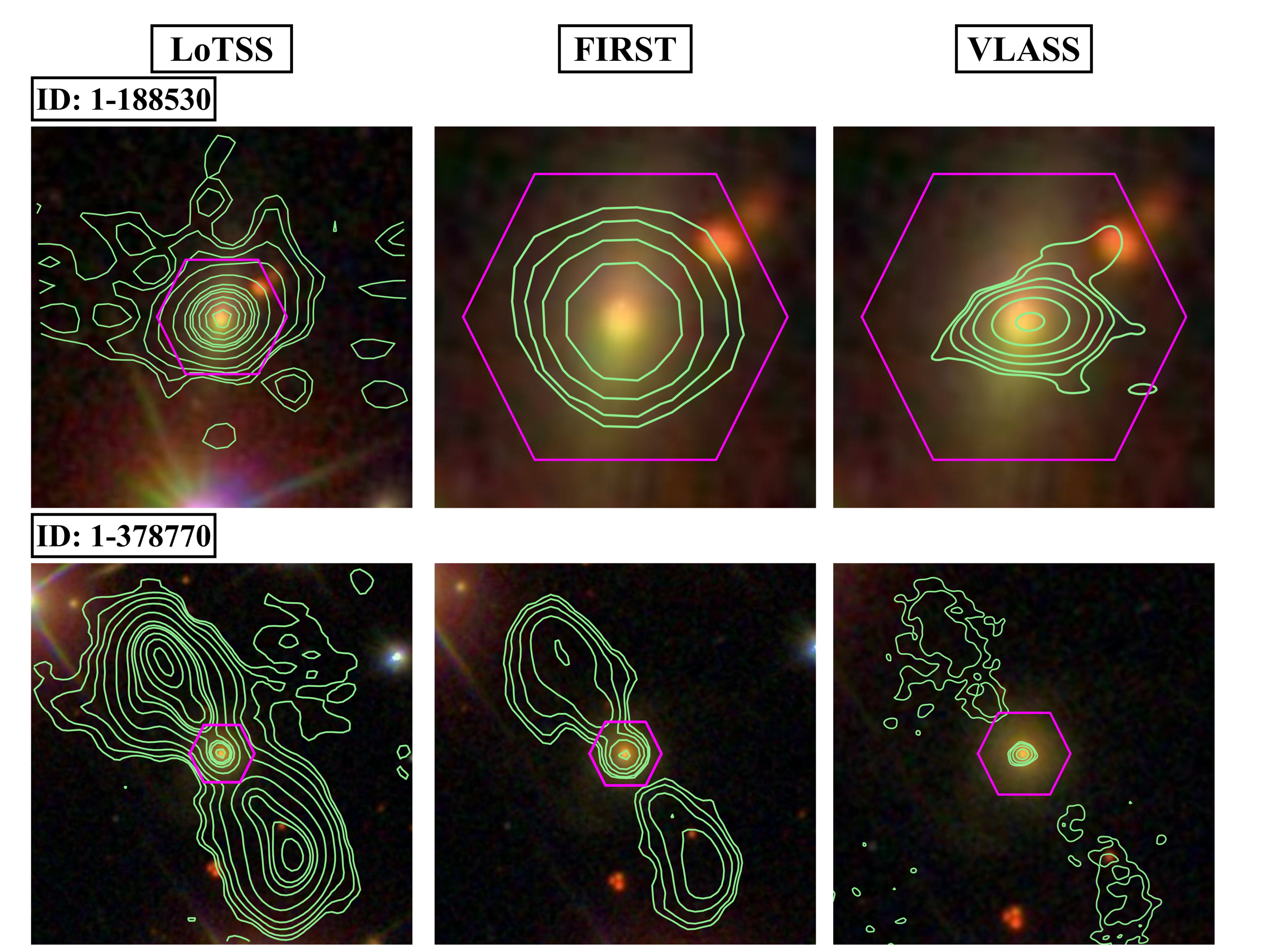}
   \caption{The LOFAR (left column), FIRST (middle column) and VLASS radio contours (right column) on top of optical images from SDSS for two red geysers with MaNGAID: 1-188530 (first row), 1-378770 (second row). The magenta hexagon signifies the MaNGA field of view. }%
   \label{compare}
    \end{figure*}

\subsection{Classification based on morphology} \label{sec:morphology}

\begin{table*} \label{tab1}
	\centering
	\caption{Summary of the radio properties for the red geyser sample. 1) MaNGA ID: identification of the galaxies. 2) z: SDSS measured redshift. 3) L$_{\rm 150MHz}$: radio luminosity measured from LoTSS survey. 4) L$_{\rm 1.4GHz}$: radio luminosity measured from FIRST survey, 5) Radio size: The largest of the linear sizes (see \S \ref{sec:size} for details) measured from LoTSS and FIRST surveys. For sources with no available LOFAR data, the FIRST measured sizes are reported as lower limits. 6) $\alpha$: Spectral index measured from radio flux from LoTSS, FIRST and VLASS surveys at frequencies of 144 MHz, 1.4 GHz and 3 GHz. 7) Morphology: The different radio morphology class from LoTSS images, as stated in \S\ref{sec:morphology}. C, D, E, I, T indicates compact, double, extended, irregular and triple class respectively. For sources with no available LOFAR data, the morphology from FIRST images are indicated in square brackets. }
	\label{tab:summary}
	\begin{tabular}{cccccccc} % four columns, alignment for each
		\hline
		\hline
		MaNGA ID & $z$ & L$_{\rm 150MHz}$ [LoTSS]  & L$_{\rm 1.4GHz}$ [FIRST]  & Radio Size  & $\alpha$ & Morphology \\
		 &  &  [$\rm 10^{22}~W~Hz^{-1}$] & [$\rm 10^{22}~W~Hz^{-1}$] & [kpc] & & \\
		\hline
		\hline
		1-575668 & 0.060 & 1.71$\pm 0.17$ &  --  &  6.24$\pm0.70$  & -- &  C \\
		1-273933 & 0.044 & 13.36$\pm 0.12$ &  2.67$\pm 0.06$  &  2.74$\pm 0.02$ & -0.712 & Unresolved \\
		1-217324 & 0.024 & 1.75$\pm 0.02$
		&  --  & 10.50$\pm 1.2$ &  -- & D\\
		1-48084 & 0.031 & 0.23$\pm 0.04$
		&  --  & 6.96$\pm 0.88$  &  -- & E \\
		1-279073 & 0.032 & 0.49$\pm 0.04$
		& 0.59$\pm 0.03$  & 3.92$\pm 0.76$ & -0.019 & C \\
		1-198182 & 0.036 & 3.35$\pm 0.06$
		& 1.41$\pm 0.04$  &  2.45$\pm 0.06$  & -0.494 & Unresolved\\
		1-44418 & 0.031 & 0.33$\pm 0.05$
		& --  &  4.16$\pm 0.72$ & -- & C \\
		1-198180 & 0.037 & 3.53$\pm 0.20$
		& --  &  21.41$\pm 02.3$ &  -- & D \\
		1-217022 & 0.024 & 0.92$\pm 0.03$
		& 0.17$\pm 0.02$  &  2.99$\pm 0.09$  & -0.616 & C \\
		1-256446  & 0.058 & 13.24$\pm 0.29$ & 2.58$\pm 0.01$   &  5.05$\pm 0.06$  & -0.748 & C\\
		1-245451  & 0.078 & 42.15$\pm 0.53$
		& 6.25$\pm 0.24$  &  8.86$\pm 0.06$   & -0.870  &  C\\
		1-256234 & 0.075  & 4.34$\pm 0.94$
		& -- &  23.96$\pm 4.43$ & --  &  E \\
		1-352569 &  0.079 & 11.07$\pm 1.58$
		& 9.51$\pm 0.23$  &   7.81$\pm 0.246$  & -0.163  & C\\
		1-322336  & 0.135 & 20.99$\pm 0.70$
		& 6.09$\pm 0.69$  &   8.51$\pm 0.35$  &   -0.496  & C\\
		1-374549 & 0.086 & 2.98$\pm 0.61$
		& --  & 9.7$\pm 2.08$  &  -- &  E \\
		1-321221 & 0.036 & 1.31$\pm 0.08$
		& 0.27$\pm 0.04$  & 3.84$\pm 0.25$   & -0.132  & C\\
		1-268789 & 0.059  & 16.53$\pm 0.29$
		&  7.04$\pm 0.12$ & 4.13$\pm 0.05$ & -0.416  & Unresolved\\
		1-595166  & 0.044 & 58.92$\pm 0.65$
		& 5.41$\pm 0.07$  & 44.09$\pm 2.7$  & -1.06 & T\\
		1-575742 & 0.061 & 3.15$\pm 0.19$
		& --  &  6.78$\pm 0.67$   &  &  C \\
		1-209772 & 0.041 & 96.92$\pm 0.16$
		& 25.78$\pm 0.07$  &  20.31$\pm 2.1$  &  -0.628 & I \\
		1-627331 & 0.027 & 4.51$\pm 0.14$
		& 0.11$\pm 0.02$  &  2.52$\pm 0.06$   &  -0.741 & C \\
		1-188530 & 0.055 & 21.24$\pm 0.23$
		& 9.06$\pm 0.11$  & 25.59$\pm 3.33$   & -0.283 & I \\
		1-605515 & 0.096 & 5.37$\pm 0.54$
		& --  & 1.2$\pm 0.09$ &  --  &  Unresolved \\
		1-150792 & 0.066 & 0.71$\pm 0.16$
		& --  & 8.41$\pm 2.04$ &  --  & C \\
		1-218116 & 0.047 & 4.86$\pm 0.12$
		& 2.23$\pm 0.08$ &  10.35$\pm 2.87$  & -0.312 & E\\
		1-634825 & 0.030 & 4.84$\pm 0.20$
		& 1.36$\pm 0.03$ &  5.36$\pm 0.06$ &  -0.306 &  C\\
		1-378770 & 0.13 & 2984.26$\pm 10.57$
		& 615.10$\pm 7.2$ &  233.07$\pm 7.2$  & -0.847 &  T\\
		1-94168 & 0.03 & 2.79$\pm 0.03$
		& 0.98$\pm 0.03$ &  20.96$\pm 1.87$  & -0.746 & D\\
		1-567948 & 0.13 & 15.08$\pm 1.00$
		& -- & 11.40$\pm 2.56$  & -- & C \\
		1-289864 & 0.049 & 695.26$\pm 15.26$
		& 8.38$\pm 0.09$ & 157.184$\pm 3.04$  & -0.900 &  T\\
		1-23958 & 0.029 & 1.27$\pm 0.03$
		& 0.34$\pm 0.03$ & 16.15$\pm 1.8$ &  -0.621 & D\\
		1-218764 & 0.068 & 3.08$\pm 0.12$
		& -- &  2.15$\pm 1.8$ &  -- & Unresolved\\
		1-584723 & 0.015 & 0.35$\pm 0.06$
		&-- & 1.59$\pm 0.1$ &  -- & Unresolved \\
		1-24104 & 0.029 & 1.56$\pm 0.07$
		& 0.39$\pm 0.03$ & 6.55$\pm 1.2$  & -0.612 & E\\
		1-113668 & 0.129 & -- & 6.39$\pm 0.06$ &  $>$0.33$\pm 0.02$ & -- & [Unresolved] \\
		1-550578 & 0.076 & -- & 8.44$\pm 0.22$ &  $>$1.68$\pm 0.62$ & -- & [C] \\
		1-37036 & 0.0283 & -- & 0.13$\pm 0.02$  & $>$ 0.58$\pm 0.7$ & -- & [Unresolved] \\
		1-43718 & 0.041 &  -- & 1.24$\pm 0.06$ &  $>$ 0.85$\pm 0.52$ & -- & [Unresolved] \\
		1-209926 & 0.095 & -- & 3.63$\pm 0.03$ & $>$ 13.04$\pm 3.5$ & -- & [E] \\
		1-210863 & 0.03 & -- & 1.02$\pm 0.06$ & $>$ 10.17$\pm 0.03$ & -- & [E]\\
		1-96290 & 0.130 & -- & 6.72$\pm 0.66$ & $>$ 11.64$\pm 3.98$ & -- & [C] \\
		1-37440 & 0.0136 & -- & 0.66$\pm 0.06$ & $>$ 0.78$\pm 0.34$ & -- & [Unresolved] \\
		\hline
		\hline

	\end{tabular}
\end{table*}
 
We primarily use data from the LoTSS survey to analyze the radio morphology of the red geysers for the following reasons. First, the LoTSS data is most sensitive to extended fainter radio emissions amongst the interferometers used in this work as it has shorter interferometric baselines than the VLA surveys. Second, the low frequency ($\sim \rm 150~MHz$) radio continuum emission reflects the oldest and the lowest energy emission from the plasma  which helps in characterizing the full extent of the structure and enables a robust classification.  
As mentioned in \S \ref{detection}, 34 out of 103 red geysers with LOFAR observations are detected at 150 MHz. Six of them are unresolved with deconvolved major axis $< \rm 5''$ (i.e. $<$ 3~kpc at median z = 0.03) and are thus physically contained within the central region of the galaxy. There are 28 LOFAR-detected red geysers which are resolved compared to the beam size. %
We visually classify the morphology of these 28 sources into five types based on the LOFAR images, roughly following \cite{baldi18, kimball11, jarvis21}: 

\begin{enumerate}

    \item Compact (C): if the source shows no visibly spatially-resolved features (i.e. has the appearance of a single two-dimensional Gaussian) and is constrained within the host galaxy. These sources have sizes larger than the beam size and may be elongated intrinsically but higher resolution is needed to confirm this.  
    
    \item Extended Jet (E): if the source is visibly spatially extended in one direction but composed of one contiguous feature, i.e., one distinct peak in the radio emission.
 
    \item  Double (D): if the source shows two distinct peaks in the radio emission. 
  
    \item Triple (T) : if the source shows three distinct peaks in the resolved radio image, typically consisting of two jets and one core.

%    ``core-jet'' structure, i.e., if the source is composed of one contiguous feature, visibly spatially extended in one direction. We note that this definition is purely to describe the morphology, and may not be physically associated with an AGN-driven jet.
    \item Irregular (I): sources generally with one distinct peak in the radio emission but with unique spatially extended (irregular) radio morphologies that do not fit within the above categories. 
%    \item Unresolved (U): sources which are unresolved and has de-convolved major axis $< 6''$. 

\end{enumerate}

The most common morphology is the ``compact'' class, found in 14 out of 28 sources. The object with MaNGAID 1-256446 in Fig.~\ref{example_lofar} belongs to this class of sources. 
    In five sources, we see a comparatively extended morphology stretching in one direction, which we define as ``extended jet''. It is to be noted that this term is used to describe the morphology only and might not be physically associated with an AGN-jet. 
    Four sources show a double-peaked radio emission (classified under ``double''). They generally consist of a radio core and a one-sided bubble/tail which may or may not be directly attached to the central component. MaNGAID 1-23958 (Fig.~\ref{example_lofar}) is one such example. 
    In three sources, we see three or more distinct peaks in radio emission. Two of them show a central core emission with large scale double-sided lobes on either side, extending $>$ 40 kpc. The third source shows an extended morphology which consist of several distinct radio peaks with $>50 \sigma$. These belong to the ``triple'' category. The galaxy with MaNGAID 1-378770 in Fig.~\ref{example_lofar} is an example with double-sided jets.  Finally, MaNGAID 1-188530 represents the irregular morphological class; in this case the LOFAR image looks quite peculiar, resembling a jelly-fish, with a bright central core superimposed on a large scale structure of diffuse radio emission.
%    Out of those 16 galaxies, the most common morphology (in 8 out of 16 sources) is almost round but slightly elongated in one direction with a spatial extent of $\rm 10 - 20$ kpc, however not extending much beyond the host galaxy (For eg., MaNGA ID: 1-273933 and 1-245451 in Fig. 5). In two sources (MaNGA ID: 1-378770 and 1-595166 in Fig. 5) we see large scale double-sided lobes with spatial extent between $60 - 80 $ kpc and another one (MaNGAID: 1-289864, not shown in Fig. 5) where the two large scale jets ( extending out to $>$ 100 kpc) are bent in the form of a downward 'V'. Four sources have a central radio core and a one-sided radio bubble/ diffuse radio structure, sometimes detached from the central component. The detached structure doesn't have any other optical counterpart and hence may or may not be associated with our central red geyser source. MaNGAID: 1-209772 (Fig 5) is one such source with a one-sided radio tail attached to the central core. Two sources (MaNGAID: ) have a core-jet shape with a central enhanced radio core and an extended low-brightness structure with the spatial extent $<$ 20 kpc although sometimes the spatial structure is quite irregular and hard to classify. Among them, one is barely resolved and of uncertain morphology. Galaxy with MaNGAID: 1-188530 (Fig 5) is quite peculiar with a jelly-fish like morphology where a bright central source superposed to a large scale structure of diffuse radio emission. 
    
     Although we perform our morphological classification primarily using LOFAR images there are eight FIRST-detected red geysers which are not covered by the LOFAR observations. We similarly categorize these sources using FIRST data into the five classes described above. Out of those eight sources, four are unresolved with a deconvolved major axis $< \rm 2.5''$. The rest belong to either the compact or extended jet classes.

    Table~\ref{tab:summary} shows the classification type of each LOFAR-detected red geyser along with the additional eight FIRST-detected sources which do not have corresponding LOFAR images (indicated by square brackets).

    Fig.~\ref{compare} shows the morphological comparison of the LoTSS images (left panel) with images from the FIRST (middle panel) and VLASS surveys (right panel) of two example red geysers out of the five sources from Fig.~\ref{example_lofar}. The radio-detected red geysers, depending on the morphological class, exhibit spatially diffuse extended features which are often only visible with LOFAR data. In most cases, the corresponding VLASS and FIRST images look rather compact. %This is primarily because of LOFAR's intrinsic good sensitivity to extended structure due to the presence of short interferometric baselines and its higher absolute sensitivity to steep-spectrum emission. 

\subsection{Radio size }  \label{sec:size}

The actual physical sizes of the radio sources are independent of the frequency of observations. However, the apparent linear size, as measured from different radio bands - i.e., using FIRST, VLASS and LoTSS data, can be different (as evident from Fig.~\ref{compare}) for several reasons. First, the FIRST/VLASS data are at higher frequencies and hence less sensitive than LOFAR to structures of typical spectral index which can lead to a smaller estimated size. Second, FIRST and VLASS lack short interferometric baselines which makes observations of extended structures above a certain size difficult in these surveys. Surface brightness sensitivity, on the other hand, can limit LOFAR sizes as well. Hence, we estimate sizes from both LoTSS and FIRST surveys, and report the larger of the two as our best estimate of the physical size of the radio source. In almost all cases, the size estimate obtained from LOFAR is greater than that from FIRST by at least a few factors.
So in cases where only FIRST data is available, we report the FIRST-measured size as a lower limit.
Our method of determination of size depends on the nature of the radio morphology.

For the sources under the label 'double' and 'triple', which show two and three distinct radio features in LOFAR images respectively, the linear size in the 150 MHz is calculated as the distance between the peak emission of the two farthest morphological features detected within $\rm 10 \ \sigma$ contours. For sources having contiguous, extended (E) and irregular morphology (I) with closely blended components, we measure the end-to-end linear size of the radio structure detected within $> \rm  10 \ \sigma$.  For the cases where the source is featureless and has only one primary morphological feature (classified as compact 'C'), we use the major axis size, de-convolved from the beam, as listed in the LOFAR catalogue. 

Depending on the specific structure of the FIRST image, the linear size in 1.4 GHz is also calculated in a similar way.
The FIRST catalogue also provides major axis measurements (FWHM in arcsec) from the elliptical Gaussian model for the source which are then deconvolved to remove blurring by the elliptical Gaussian point-spread function.  As mentioned above, we consider the larger of the measured sizes from LOFAR and FIRST to be the ``largest linear size'' or simply the radio size.
Table~\ref{tab:summary} lists the measured sizes of the radio sources along with 1$\sigma$ uncertainty. For the compact and unresolved sources, the errors are derived from the respective catalogue which reports 1$\sigma$ uncertainty in the de-convolved major axes, derived from the Gaussian models of the sources. For resolved objects showing spatially extended morphologies, we assume the uncertainty to be the linear size (in kpc) corresponding to half the beam-width, i.e. 3$''$ for LOFAR. The uncertainty in size from the FIRST band  is obtained from the relation:
$\rm \sigma_{Size} = 10'' * (1/SNR + 1/75)$ \footnote{http://sundog.stsci.edu/first/catalogs/readme.html}, where SNR is the signal-to-noise, given by: $ \rm SNR = (F_{peak}-0.25) / RMS.$ $\rm F_{peak}$ and RMS signifies the peak flux and the root mean square deviation measured from the catalogue respectively.

\begin{figure*}
   \centering
   \graphicspath{{./plots/}}
   \includegraphics[width = \textwidth]{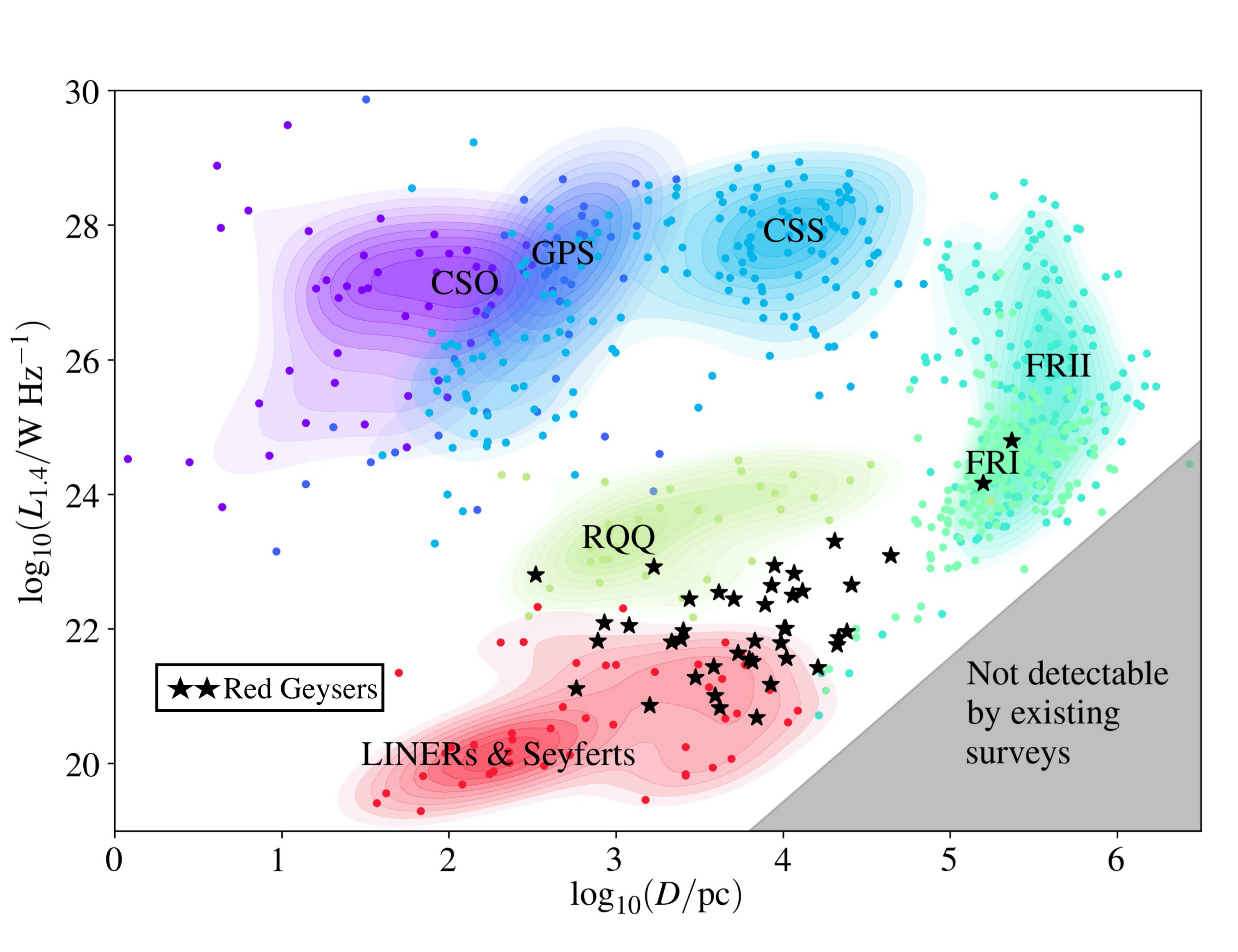}
   \caption{Radio luminosity versus linear size for our red geyser sample (black stars) compared to a sample of radio AGN population compiled from the literature by \cite{jarvis19} and \cite{hardcastle20}. Points show individual objects and the colored contours represent source density. Seyfert and LINER galaxies are from \cite{gallimore06} and \cite{baldi18}, while the radio-quiet quasars (RQQ) are from \cite{jarvis19} and \cite{kukula98}. The rest of the objects, consisting of compact steep spectrum (CSS) sources, gigahertz peaked spectrum (GPS), compact symmetric objects (CSO) and Fanaroff-Riley class 1 and 2 are categorized by \cite{an12, mingo19}. %
   \label{power_size}}
    \end{figure*}

Table~\ref{tab:summary} shows the radio sizes along with radio luminosities of the radio detected red geysers in both the LOFAR and FIRST radio bands. Fig.~\ref{power_size} shows the radio-luminosity (at 1.4 GHz) vs. linear size of the radio-detected red geysers, detected in at least one of the radio bands (in black stars), over-plotted along with different classes of radio sources from the literature in different colored contours. For the red geysers with available LoTSS counterpart, we convert measured 144 MHz flux from LOFAR to 1.4 GHz, assuming a spectral index of -0.7 in order to calculate the desired luminosity \citep{condon02}. This is preferred over using FIRST flux that misses extended emission. However for those without LOFAR data, corresponding FIRST luminosities are used. The different colored contours in the figure, as indicated, represent compact symmetric objects (CSO), gigahertz peaked spectrum (GPS), compact steep spectrum sources (CSS), Fanaroff-Riley class 1 (FRI), Fanaroff-Riley class 2 (FRII), radio-quiet quasars (RQQ) and Seyferts. The data for these radio-detected AGNs have been compiled by \cite{jarvis19} from a variety of studies of radio AGN population, namely \cite{an12, gallimore06, kukula98, baldi18, mingo19}. Most of the radio-detected red geyser sources, marked as black stars, overlap with the LINER/Seyfert type classification and with the tail of the distribution of radio-quiet quasars. There are two sources which lie on the FRI part of the diagram, both of which are categorized as ``triple'' according to morphology classification. Our sources are in general consistent with having similar small scale low-luminosity jetted morphologies as observed in the radio-quiet quasars in \cite{jarvis19} or with FR0 sources which remain classified as ``compact'' unless higher resolution observations are available to resolve the sources, as in \cite{baldi15, capetti19, hardcastle19}.

\subsection{Spectral Index} \label{spectral_index}

   \begin{figure}[h!!]
   \centering
   \includegraphics[width = 0.5\textwidth]{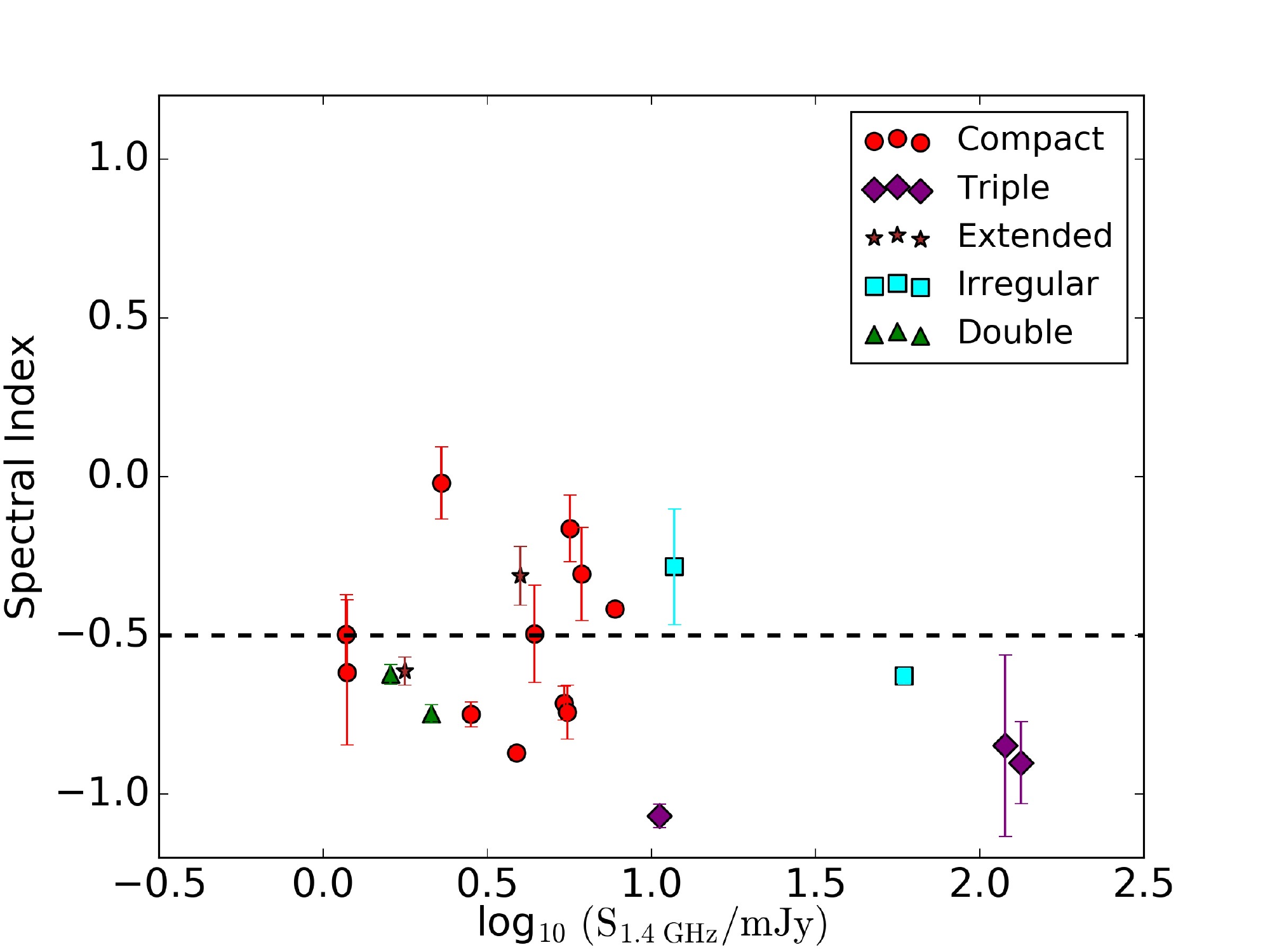}
   \caption{The spectral index vs. radio flux density at 1.4 GHz frequency for the 21 red geysers which are simultaneously detected in LOFAR, FIRST and VLASS. The galaxies are color-coded by radio morphology classification, as defined in \S 5.1. Note, for this figure, the unresolved galaxies are also marked as 
  ``compact'' along with the actual compact class.}
  \label{fig:spectral}
    \end{figure}
     
In this section, we study the spectral indices for our sample of 21 red geysers which have simultaneous radio detection from LOFAR, FIRST and VLASS data (See \S\ref{detection} for more details). %Relationships between integrated spectral properties, radio power and size have been seen in large samples in previous studies (e.g. de Gasperin et al. 2018; Tisanić et al. 2020). 
    The radio-continuum spectrum is generally dominated by non-thermal synchrotron emission with the characteristic power-law, S$_{\rm \nu} \propto \rm \nu^{\alpha}$, where $\alpha$ is the spectral index, $\nu$ is the frequency of radio emission and S$_{\nu}$ is the flux density measured at frequency $\nu$. In star-forming galaxies there may also be some additional contribution from the thermal bremsstrahlung (free-free) emission \citep{duric98, gioia82}, but that is irrelevant here.

The spectral index of a radio galaxy can provide information about the relative contributions of the core and the extended lobed structure in the total radio emission. Core-dominated emission and any compact source typically have flat ($\rm \alpha > - 0.5 $) spectrum due to the effects of synchrotron self absorption and free-free absorption \citep[e.g.,][]{odea21}. The extended lobes, on the other hand, tend to have steep spectra ($\alpha \leq -0.5$) because the predominant emission mechanism is optically-thin synchrotron. Thus, radio sources which are more core-dominated therefore tend to have flatter spectra than those dominated by extended emission. This is confirmed from the observation of a high-core dominance in FR0s, owing to their compact nature, based on high resolution images \citep{baldi19}.
     
     As can be seen in Fig.~\ref{example_lofar} and Table~\ref{tab:summary}, our sources exhibit a range of radio morphologies from LOFAR data. %Hence accurate estimation of total flux densities is critical to get precise spectral index measurements. 
     We use the integrated flux densities and uncertainties from the LoTSS, FIRST and VLASS catalogs for our analyses (See Table~\ref{tab:summary} for the flux densities and uncertainties for each source). For sources having more than one component/ region, we visually identify all the individual components detected at a significance of $> 5 \sigma  $ which are associated with the source and add the flux densities from these components together.  
     We define the radio spectral index, $\alpha$, using S$_{\rm \nu} \propto \nu^{\alpha}$ and measure $\alpha$ by fitting the flux densities measured at three different frequencies (1.4 GHz, 3 GHz and 150 MHz) with the said function for each source. The errors are obtained from the uncertainties in flux densities via simple error propagation. Table~\ref{tab:summary} lists the spectral indices with 1$\rm \sigma$ uncertainty of the 21 red geyser galaxies.  
     
     The red geysers show a large spread in spectral indices, ranging from $-1.0$ to 0.0 with a median value of $-0.62$ (see Table.~\ref{tab:summary} and Fig.~\ref{fig:spectral}). The extended radio sources, which do not belong to the “compact” morphological class, exhibit steeper spectra on average with a mean spectral index $\alpha = -0.67$. The fraction of sources with a steep spectrum ($\alpha < -0.5$) is $\sim \rm 57$\% indicating the presence of extended emission or dominance of lobed components in those radio sources. This automatically implies that 43\% of our sample of 21 red geysers (i.e., eight sources) have a flatter spectrum, with core-dominated compact structures. This agrees well with our morphological classifications, in that a fairly large fraction of red geysers have compact radio morphologies confined within the host galaxy. The bright nuclear radio component with a moderately flat spectral index (i.e.,$\alpha > -0.5$) may indicate a contribution from radio emission associated directly with an AGN ‘core’ / unresolved base of the jet \citep{padovani16}. 
     
     Fig.~\ref{fig:spectral} shows the variation of the spectral index with 1.4 GHz flux density as measured from FIRST. Galaxies are color-coded by the morphological class, as given in Table~\ref{tab:summary}. Note, for this figure, the unresolved sources are also color-coded as the ``compact'' class. 
     All three galaxies classified as ``triple'' have steep spectral indices indicating that those sources have more extended lobes with predominantly optically thin synchrotron emission. With an exception of two galaxies, almost all the red geysers which are not-compact (any color except red) have a steep spectrum and lie below the dashed line. The two sources showing flat spectrum might be due to the result of a combination of underlying biases in the measured fluxes and puzzling morphology leading to an incorrect classification. On the other hand, half of the sources labelled ``compact'' show flat spectrum while the other half exhibit a steep spectral index. 
     %integrated radio spectrum of radio jets is typically associated with a flat spectral slope due to the superposition of self-absorbed synchrotron spectra. 
     We do not see any significant correlation of spectral indices with radio flux density. The implications of these results are discussed in \S \ref{discussion}.

\section{Radio jet and connection with galactic outflows} \label{sec:ionized}

   \begin{figure}[h!!]
   \centering
   \includegraphics[width = 0.5\textwidth]{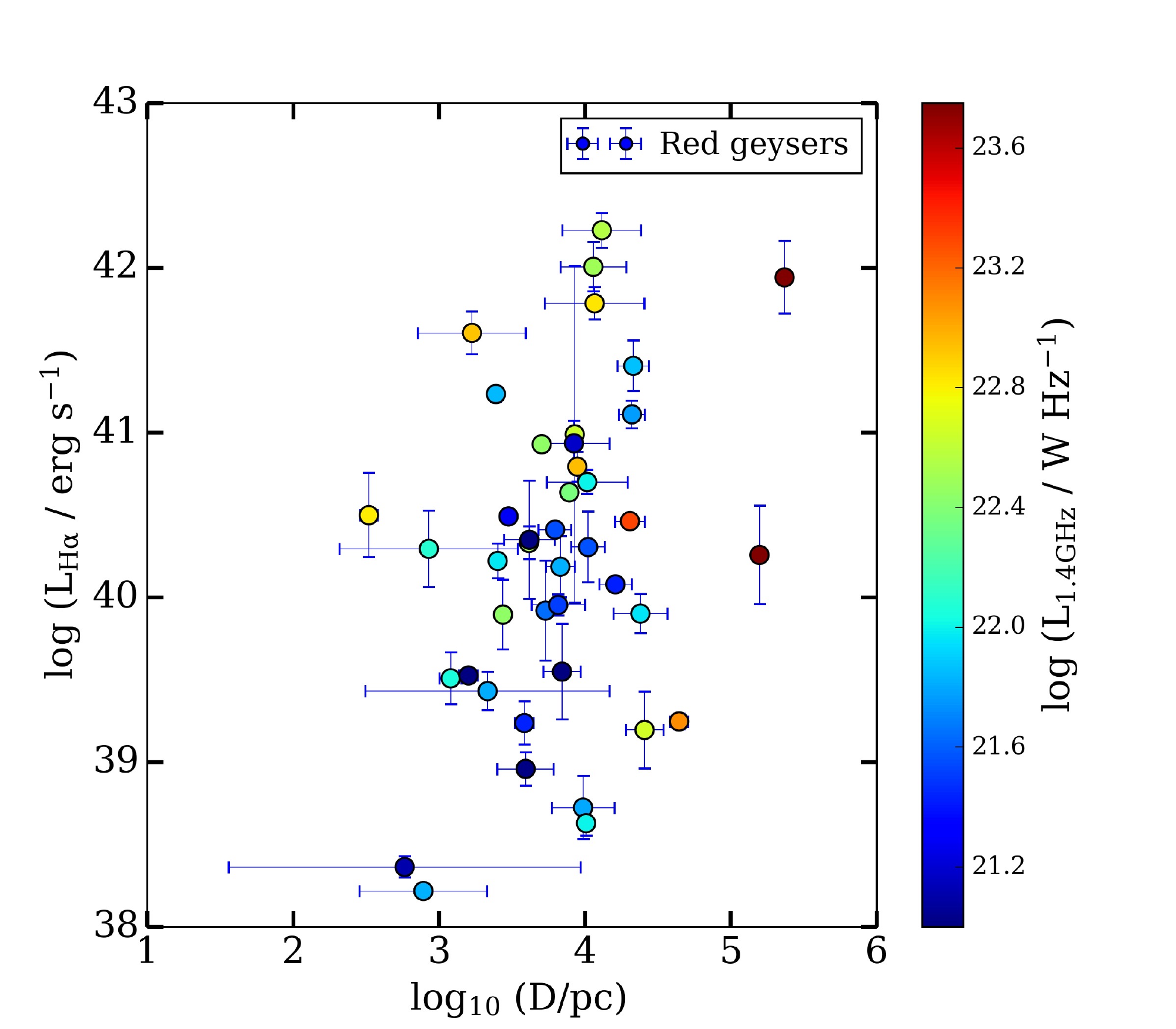}
   \caption{ H$\alpha$ luminosity integrated over one effective radius aperture vs. radio size (as given in Table~\ref{tab:summary}) for 42 radio detected red geysers, color coded by radio luminosity at 1.4 GHz. For the 34 LOFAR detected sources, L$_{\rm 150MHz}$ has been converted to L$_{\rm 1.4GHz}$ assuming a spectral index of -0.7. For the remaining eight galaxies outside LOFAR footprint, FIRST measured luminosities have been used. A moderate positive correlation between H$\alpha$ luminosities and radio size is detected.  }%
    \label{halpha_size}
    \end{figure}

  \begin{figure}[h!!]
   \centering
   \includegraphics[width = 0.5\textwidth]{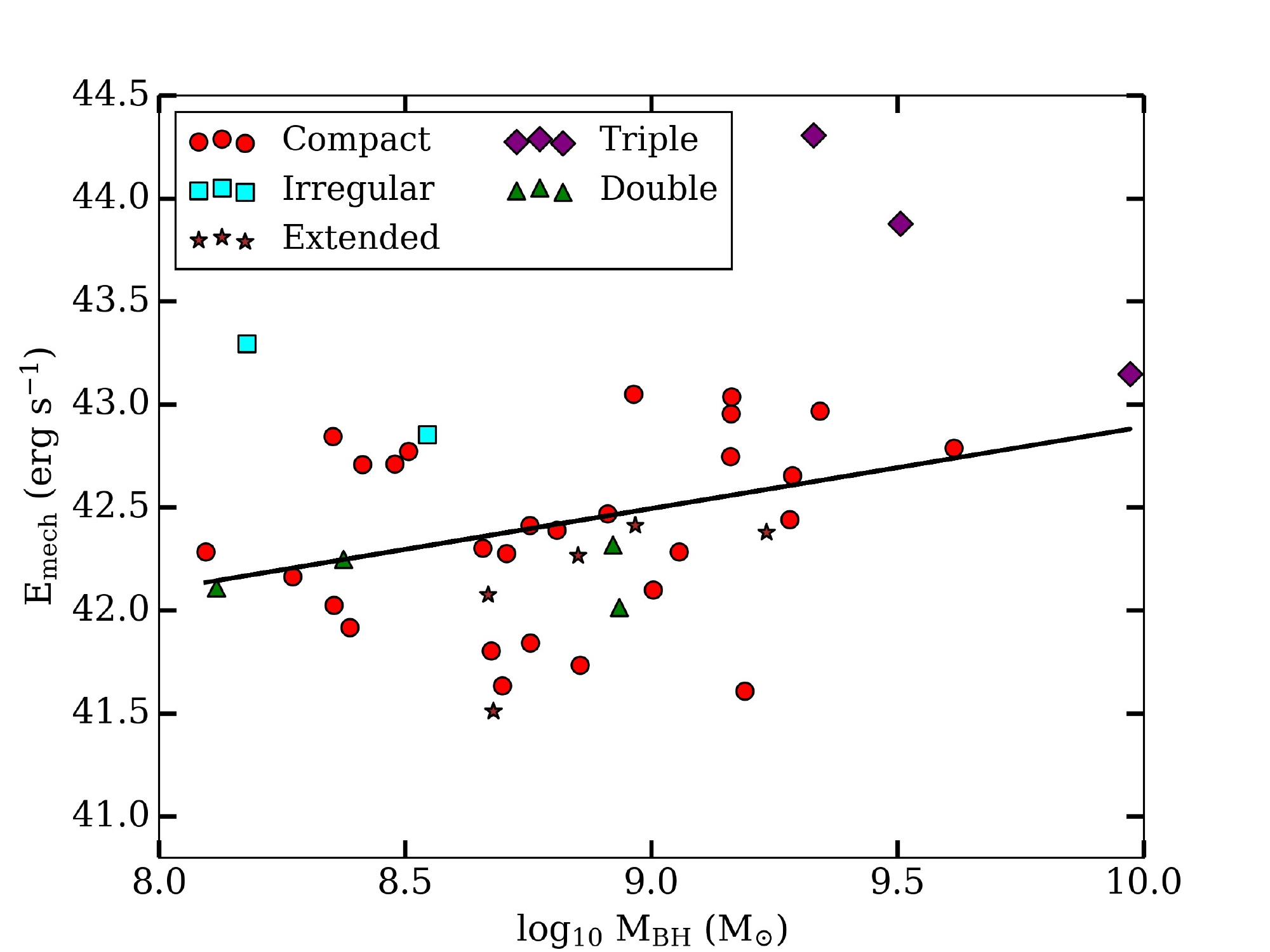}
   \caption{ Jet mechanical energy, estimated from radio luminosity vs. supermassive black hole mass for 42 radio detected red geysers color coded by morphological class. A clear positive trend is seen, implying that the observed radio emission is linked to the nuclear SMBH.The black solid line shows the best fit relation obtained by least square optimization. }%
    \label{jet_smbh}
    \end{figure}

       \begin{figure}[h!!]
   \centering
   \includegraphics[width = 0.5\textwidth]{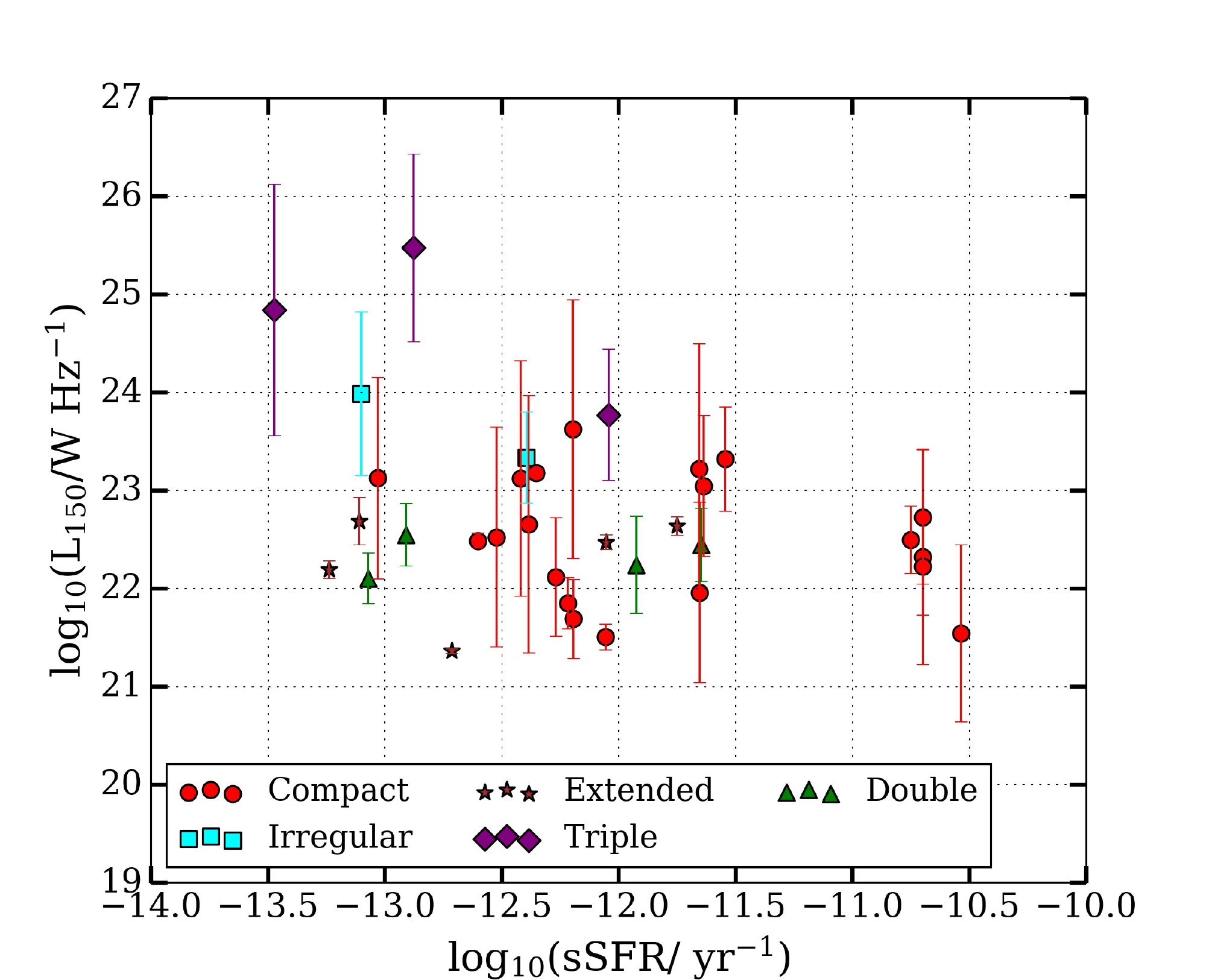}
   \caption{ Radio luminosity (L$_{\rm 150MHz}$) vs. specific star formation rate for the radio detected red geysers, color coded by morphological class. Galaxies which are not classified as ``compact'' and possess extended radio structures show low SFR for their stellar mass. This implies large radio sources are more efficient in quenching red geyser galaxies.  }
    \label{ssfr_l150}
    \end{figure}

%   \begin{figure}[h!!]
%   \centering
%   \includegraphics[width = 0.5\textwidth]{ha_size.jpeg}
%   \caption{Average H$\alpha$ luminosity (from MaNGA single Gaussian fits to emission lines across 1Re) versus largest
%linear radio size of the radio emission in the radio-detected red geysers, as observed by LOFAR in 144 MHz frequency band. There is a hint of a positive correlation with high L$_{H\alpha}$ sources preferentially associated with larger radio sources. }%
%    \label{halpha_size}
%    \end{figure}

       \begin{figure*}
   \centering
   \includegraphics[width = \textwidth]{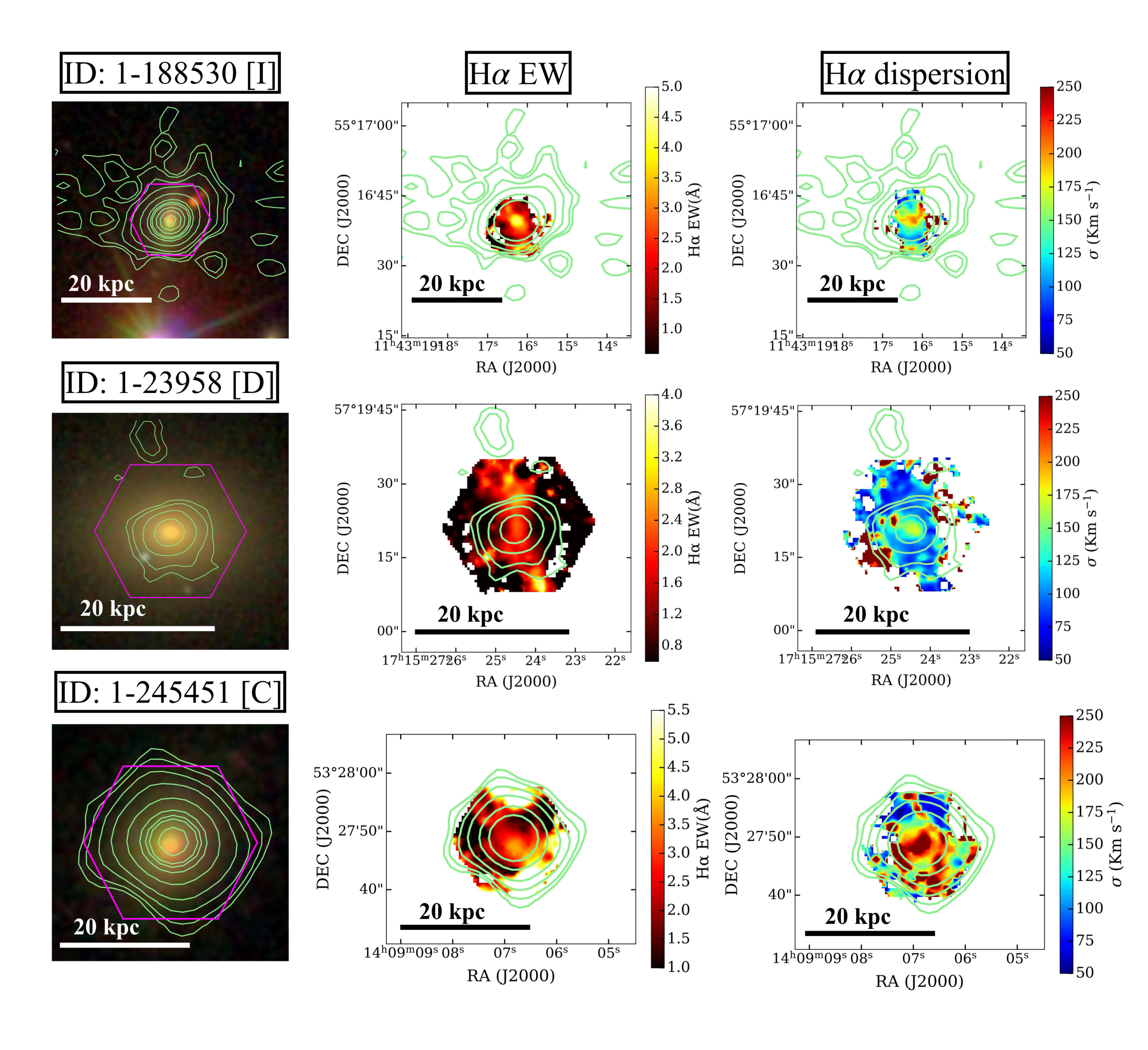}
   \caption{ SDSS optical image (left column), MaNGA H$\alpha$ EW map (middle column) and H$\alpha$ velocity dispersion (right column) shown for three red geyser galaxies with LOFAR radio contours overplotted in green. Each row represent a red geyser galaxy. Galaxy with ID: 1-188530 (top row) belongs to the irregular morphology class, 1-23958 (middle row) has a double morphology and 1-245451 (bottom row) shows compact radio structure.   }%
    \label{fig:ionized}
    \end{figure*}

      \begin{figure*}
   \centering
   \includegraphics[width = \textwidth]{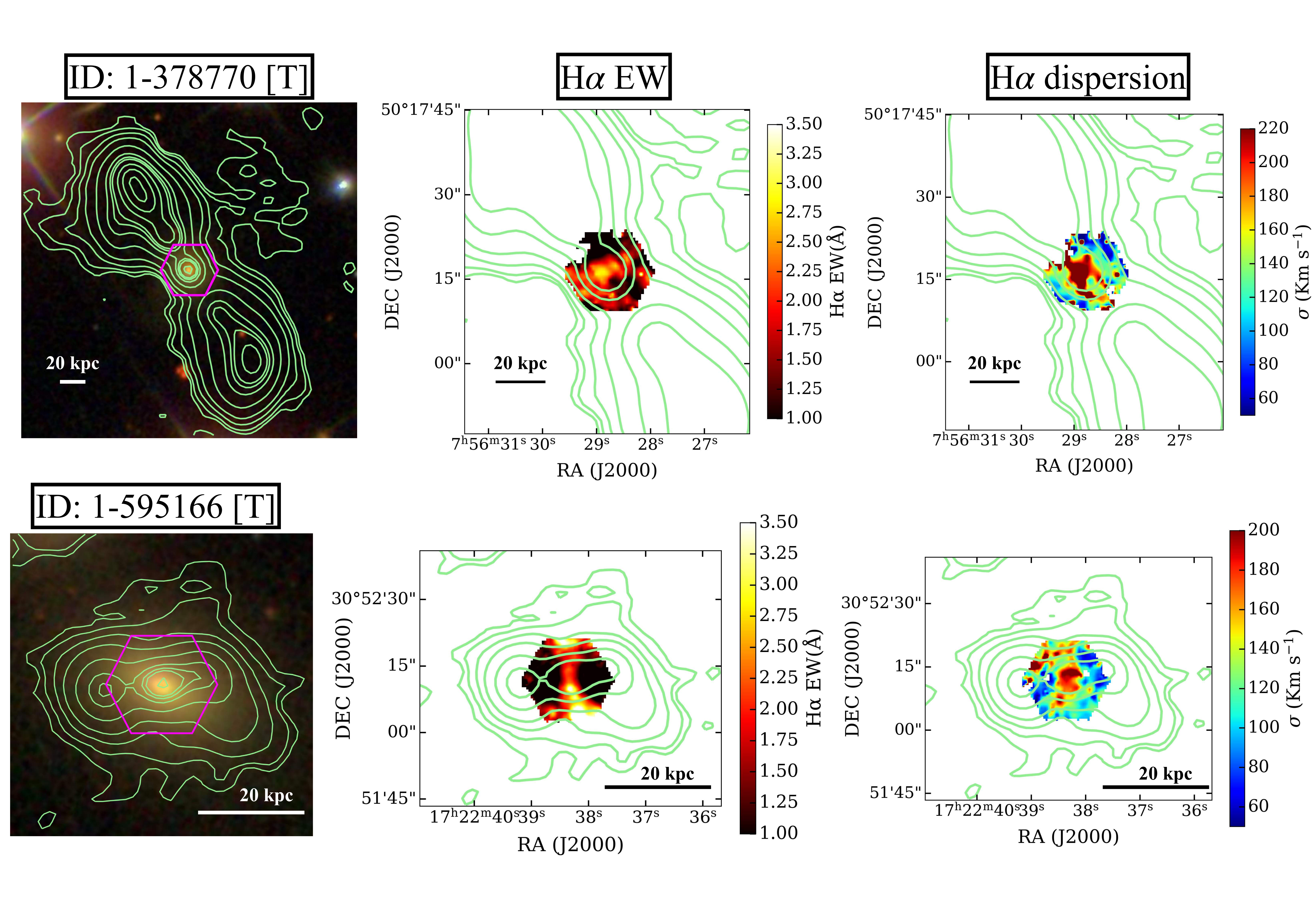}
   \caption{ SDSS optical image (left column), MaNGA H$\alpha$ EW
          map (middle column) and H$\alpha$ velocity dispersion (right
          column) shown for two red geyser galaxies with LOFAR radio
          contours overplotted in green. Each row represent a red
          geyser galaxy. Both galaxies belong to the ``triple'' morphology class showing large scale radio lobes.   }%
    \label{fig:ionized2}
    \end{figure*}

%%%%%%%%%%%%%%%%%%%%%

    %% include Ha luminosity vs radio linear size plot?
    
    % include radio size vs. Ha FWHM (dispersion) plot. 

    Here we explore the connection
between the radio emission, its morphology, and the ionized gas properties in our sample. %and compare this to similar radio AGN samples in  the literature. %We compare integrated H$\alpha$ luminosity and line widths, split by morphological classification and then proceed to investigate the spatially resolved maps on a case by case basis. 

\subsection{Integrated properties} \label{ionized_1}

Traditionally, for compact radio-loud sources (CSS/GPS),
a weak positive trend has been observed between the luminosity
of strong emission lines and radio size \citep{odea21}. Larger radio sources are found to be more commonly
associated with higher [OIII] luminosities \citep{odea98}.
A similar correlation has also been observed in radio-quiet
sources using SDSS-measured [OIII] luminosities \citep{jarvis21}. Here we investigate whether similar correlation exists in red geyser galaxies. Here we use H$\alpha$ as a tracer of ionized gas instead of the traditional [OIII] line, since the characteristic bi-symmetric pattern identifying a red geyser is most prominently observed in the spatial distribution of H$\alpha$.%In order to investigate the inter-relationship between gas velocity dispersion, luminosity of emission lines and the radio luminosities in red geysers, 
%Larger radio sources are found to be more commonly associated with higher [OIII] luminosities \citep{odea98}. A similar correlation has also been observed in radio-quiet sources using SDSS-measured [OIII] luminosities \citep{jarvis21}. Here we investigate whether similar correlation exists in red geyser galaxies which are passive early type galaxies that exhibit a compact, radio-quiet nature.

Fig.~\ref{halpha_size} shows the the H$\alpha$ luminosity, integrated over one effective radius as observed by MaNGA, vs. the linear size of the radio emission of the 42 radio detected red geysers. The data are color-coded by the radio luminosity at 1.4 GHz. The radio luminosities are derived by converting L$_{\rm 150MHz}$ to L$_{\rm 1.4GHz}$ assuming a spectral index of -0.7, as before. Our data show a positive correlation that is consistent with what has been observed in radio-loud compact sources and in some recently studied radio-quiet sources. \cite{liao20} showed that for the radio-loud population, the
[OIII] bright sources (with L[OIII] $\rm > 10^{42}~erg~s^{-1}$) have radio sizes $>$ 0.75 kpc. This is consistent with our result since the red geysers show radio sizes $>$ 10 kpc for similar L$\rm _{H\alpha}$. The existence of this positive correlation in the red geysers is strongly supported by a Spearman rank correlation coefficient of 0.6. Fig.~\ref{halpha_size} also shows that the radio luminosity generally shows a higher value ($\sim \rm  10^{23}~W~Hz^{-1}$) for larger radio size and higher H$\alpha$ luminosities (i.e. towards the upper right portion of the figure). On the other hand, as L$\rm _{1.4 GHz}$ drops to $\rm 10^{21.5}~W~Hz^{-1}$ for L$\rm _{[OIII]} < 10^{40}~erg~s^{-1} $ the radio size decreases to $< \rm 5 kpc$. This is expected because of the first order dependence of the morphology of the radio sources with radio power. Large scale radio structures are found to be generally more abundant in radio-loud sources, although this correlation is not so apparent in radio quiet sources \citep{morganti20}. In the case of red geysers, this possibly implies that larger radio sources with greater radio luminosity hosts greater amount of ionized gas.

In previous studies, similar relations between the strength of an ionized gas outflow tracer and the size of the radio source has been explained by considering the interaction between an embedded, AGN-driven radio jet and the ISM. 
% Regardless of the morphological classification, the sizes and structures of the radio jets - which might not be fully resolved owing to spatial resolution, the important parameter to quantify the impact on the surrounding ISM is jet energy. 
Regardless of size and structure of the radio jet---which is unlikely to be resolved owing to spatial resolution in this case---the important parameter to quantify the impact on the surrounding ISM is jet energy. Although there have been several methods to estimate the jet power, the method presented by \cite{williot99} based on the synchrotron properties of the radio sources has been shown to be particularly effective.
%In order to re-affirm the association of the radio emission with the central radio mode AGN, Fig.~\ref{jet_smbh} shows the relation between jet mechanical energy and the black hole mass. 
They proposed the following conversion between the jet mechanical energy and the radio luminosity:

\begin{equation} \label{e_mech}
    \rm E_{mech} = 2.8 \times 10^{37} \left( \frac {L_{1.4 GHz}}{10^{25}\ W \ Hz^{-1}} \right)^{0.68} \ W
\end{equation}

\noindent This expression is used in this work, although there are limitations of this approach, as discussed in \cite{hardcastle13, croston18}. 
We find that red geysers show a large range in $\rm E_{mech}$, spanning three orders of magnitude, within $\rm 10^{41.5} - 10^{44.5}\ erg\ s^{-1}$. In order to re-affirm the hypothesis that the source of this estimated energy is the central radio AGN, we also calculate the corresponding supermassive black hole mass (SMBH) of our sources. The SMBH mass and the jet mechanical energy are typically connected in an AGN. A radio AGN tends to be more radio loud and is associated with more energy as the black hole becomes
more massive \citep{best05}. However, these two quantities are not expected to show any correlation for radio sources associated with other astrophysical phenomenon. We estimate the black
hole mass, M$_{\rm BH}$, using the following relation \citep{mcconnell13, cheung16}:
\begin{equation} \label{m_bh}
    \rm log_{10} \ (M_{BH}/M_{\odot}) = 8.32 + 5.64\ log_{10} [\sigma_{\star}/(200\  km s^{-1})]
\end{equation}

\noindent where $\rm \sigma_{\star}$ is the velocity dispersion of the stars, extracted from the central 2$''$ radius aperture. 
Fig.~\ref{jet_smbh} shows the the jet mechanical energy vs. SMBH mass in the radio detected red geysers, color coded by their respective morphological class. We find that they are moderately correlated. We fit the data points with a linear function using least square optimization technique and find a relation in the form:
$\rm log_{10}\ (E_{mech}) = [log_{10} \ (M_{BH})]^{0.39} + 38.925$, with a Spearman's correlation coefficient (r-value) of 0.4.
This positive trend implies that the central AGN, with possibly unresolved small-scale radio jets, is driving the radio emission seen in these galaxies.

In order to estimate the possible contribution of this jet energy to the quenching of star formation, Fig.~\ref{ssfr_l150} shows the radio luminosity at 150 MHz vs specific star formation rate (sSFR) for the LOFAR detected sources, color coded by their morphological classification. Similar to Fig.~\ref{fig:spectral}, the unresolved sources are also indicated as ``compact''. We find that the radio sources which are non-compact and belongs to either one of extended, irregular, double or triple class, have much lower sSFR with average log$_{10}$ sSFR = -13~yr$^{-1}$, compared to the compact sources with average log$_{10}$ sSFR  = -11.75~yr$^{-1}$. This implies that the radio sources showing more extended radio morphology are either more effective in quenching or reside in a larger halo with greater stellar mass, bringing the total sSFR down by several factors. A weak negative correlation is also visible between radio luminosity and sSFR, although this apparent trend can be due to the low sample size and driven predominantly by a few large and extended radio sources which are generally more radio-loud 
on average. This seems to be the case here since the three sources under the ``triple'' category and one ``irregular'' source, which are solely responsible for driving the negative trend, have radio luminosity ($\rm L_{150MHz}$) at least an order of magnitude more than the average luminosity of the rest of the sources. 
%Hence, this apparent negative trend might simply arise because of the low sample size. 

\subsection{Spatially resolved properties}

In addition to the integrated properties discussed above, we now compare the spatially resolved ionized gas flux and kinematic maps with radio image morphology.

% have enabled us to study the interaction between the ionized outflow and the radio AGN directly on a case by case basis. %, we can now study the properties of the jet-ISM interaction directly. 
% In addition to the integrated properties discussed above, the access to spatially resolved ionized gas flux and kinematic maps along with radio images have enabled us to study the interaction between the ionized outflow and the radio AGN directly on a case by case basis. %, we can now study the properties of the jet-ISM interaction directly. 
% A direct quantitative comparison between the radio and optical maps is difficult, owing to small samples and the low spatial resolution (5$''$) of the radio data. Nevertheless, we search for some visible trends from the resolved maps to understand the underlying mechanism in these sources. 

While every galaxy in the red geyser sample shows signatures of ionized gas outflows via extended bi-symmetric pattern in equivalent width map, there is a distinct lack of extended visible radio lobes on a similar scale in these galaxies as observed from a combination of LoTSS, FIRST and VLASS survey. As reported in \S \ref{sec:morphology}, 14 out of 28 ($\sim 50$\%) sources showing resolved radio emission in LOFAR observations display a compact radio morphology. Although similar radio-quiet compact sources hosting radio AGN have been observed to host small-scale ($\sim$ 1 kpc) radio jets in higher spatial resolution ($< 1''$) radio observations in previous studies \citep{jarvis19, jarvis21, venturi21,panessa19,baldi18, webster21}, the presence of resolved radio jets in the red geyser sample is observed to be quite rare with the current 5$''$ resolution. 

For the red geysers belonging to the ``compact'' or ``extended'' class, the spatial correlation between ionized wind cone and the radio jets is difficult to infer. However, the sources belonging to the ``double'' or ``triple'' category provide the most insight. In two out of three red geysers belonging to the ``triple'' class (MaNGAID: 1-378770 and 1-595166), the  radio lobes align  perpendicular to the direction of the ionized wind cone. On the other hand, for the rest of the sources which are resolved but  not  ``compact'' (four ``double'', one ``triple'' and two ``irregular''), the elongation axis in the radio images roughly aligns with the direction of the ionized wind cone.

We choose five prototypical galaxies representing the compact, irregular, double and triple morphological class to explore the detailed ionized gas-radio interaction using spatially resolved maps.

MaNGAID: 1-245451 in Fig.~\ref{fig:ionized} shows an example of a radio detected red geyser belonging to the ``compact'' radio morphology class. The three columns in the figure correspond to SDSS optical image (first row), H$\alpha$-EW (second row) and velocity dispersion (third row) extracted from the H$\alpha$ emission line. The LOFAR radio contours (in green) are over-plotted on top of each map. The on-sky diameter of the MaNGA fiber bundles (overplotted in magenta hexagon in the optical image) generally ranges between 17$''$ - 32$''$, corresponding to a physical size of $\rm 10 - 30 \ kpc$ at median redshift of MaNGA observations ($z \sim 0.03$). The bi-symmetric pattern in the H$\alpha$ EW map traces the ionized wind cone. The absence of structures in the radio image makes it hard to associate the radio properties with any specific ionized features. However, we note very high gas dispersion within the inner radio contours detected with $>20 \sigma$, indicating extreme ionized outflow kinematics there.

    For the red geyser with MaNGA ID $-$ 1-188530 (belonging to the ``irregular'' morphology class), the LOFAR image has an unusual extended morphology spanning a distance of $\sim 30$ kpc, with a central radio core and a plateau of diffuse emission. Interestingly, the radio structure is spatially extended in the direction of the H$\alpha$ enhancement in the EW map, similar to the other sources in the "irregular" class. Additionally, we note that the galaxy shows an elongated region of enhanced line width ($> 200$ km s$^{-1}$), spanning about $>$7 kpc. 
    
    In red geyser with MaNGA ID: 1-23958, the radio image show a one-sided low surface brightness bubble, detached from the central bright core. Similar to the previous example, the direction of the radio bubble roughly aligns with the bi-symmetric pattern in the $\Ha$ emission line map. This is the case for all sources classified as "double". 
    
    Finally, as already mentioned, double-lobes and distinct jets are observed to be quite rare in the red geyser sample using the current $\sim 5''$ resolution of the radio images. Out of the three sources classified in the ``triple'' morphological category, two of them (Fig 12, MaNGA ID: 1-378770 and 1-595166) have radio jets lying perpendicular to the ionized gas traced by the H$\alpha$ EW map (Fig.~\ref{fig:ionized2}). This is unlike what is observed in the above cases. The gas velocity dispersion is enhanced perpendicular to the H$\alpha$ bi-symmetric feature in 1-378770, but is aligned in 1-595166. The implications of these findings are discussed in more detail in \S \ref{discussion}.

\section{Discussion} \label{discussion}

 We have presented 150 MHz, 1.4 GHz and 3 GHz radio imaging from the LoTSS, FIRST and VLASS surveys, together with spatially resolved optical spectroscopy from the SDSS IV- MaNGA survey, for 42 radio detected red geyser sub-sample out of the total 140 $z < 0.1$ red geyser galaxies. 103 out of those 140 galaxies have available LOFAR imaging data with 34 of them ($\sim 33$\%) being radio-detected. 29 out of 140 ($\sim 21$\%) galaxies are detected in FIRST and 29 are detected in VLASS. There are 21 sources ($\sim 15$\%) which have simultaneous radio detection from all three surveys and 42 sources which are detected in at least one of them. %In this section, we aim to discuss about the origin and morphology of the radio emission from these radio-detected sources, the relation with their ionized gas and its connection in the context of radio AGN-driven wind population.  
 The radio properties are summarized in table~\ref{tab:summary}. 
 
 %%% Include our sample in context of other similar studies from Jarvis et al. 2021, also compare to radio-loud samples (Jarvis et al. discussion) and FR0 and radio quiet quasars.  
 % Include estimate of bolometric luminosities from OIII luminosities. Also eddington scale accretion rate from radio luminosities. 

 %% talk about the detection rate. why lofar detection is more. FIRST detection consistent with roy et al. control sample has lesser detection. host galaxy properties specially stellar mass, luminosity 

  The FIRST detection rate is roughly in agreement with our previous work \citep{roy18} which established that the red geysers, that show signatures of kpc-scale winds in warm ionized gas tracers, have a higher incidence of radio continuum emission than typical early type galaxies without such signatures. Thus, H$\alpha$, one of the primary tracers of putative ionized  winds in the red geysers, was seen to be associated with increased radio emission. Indeed, even within the red geyser sample, we have shown in Fig.~\ref{host} (upper right) that a radio-detection indicates a greater amount of ionized gas. The mean of the distribution of luminosity of H$\alpha$ (L$_{\rm H\alpha}$) in radio detected red geyser sample is $\rm 10^{40.5}~erg~s^{-1}$, about four-five times higher than the non-radio detected sample.

   39 out of 42 radio-detected galaxies in our sample are classified as being ‘radio-quiet’ based
  on standard criteria of \cite{xu99} (Fig.~\ref{radio_quiet}), while 38 out of 42 sources are radio-quiet from R parameter value  \citep{ivezic02}. R values in the three radio loud cases are fairly moderate, with R$<$1.5, compared to the typical R value of 2.8 for radio-loud sources from \cite{ivezic02}. Thus, the red geyser galaxies are largely low-luminosity sources and belongs to the ``radio-quiet'' group of objects \cite[see][for detailed discussions]{kellermann89, morganti20}. 
  
  \subsection{Origin of the detected radio emission} \label{discussion_1}
  
  Although the radio emission in ``radio-quiet'' sources is often attributed to being dominated by star formation processes, red geyser targets have very little star formation activity \citep[log SFR $< \rm 10^{-2} M_{\odot}~yr^{-1}$][ and Fig.~\ref{host}]{roy21}. \S \ref{fig:quenched} shows the quiescent nature of these galaxies via WISE infra red colors, confirms complete lack of star formation and presence of old stellar population through the 'D4000 vs. L$_{\rm rad}$/M$_{\star}$' method and show the lack of ionization from young stars via the BPT diagram. Indeed, in \cite{cheung16}, the central radio continuum emission in the prototypical red geyser was from a low-luminosity radio AGN ($\rm L_{1.4 GHz} \sim 10^{21}~W~Hz^{-1}$) with low Eddington ratio ($\rm \lambda \sim 10^{-4} $). \cite{roy18} showed that the expected SFR ($\sim \rm 1~M_{\odot}~yr^{-1}$) derived from the average radio luminosity from the red geysers sample exceeds the observed SFR, derived from ultra-violet to infrared SED fitting, by two to three orders of magnitude. 
  If we perform a similar calculation on our current sample of 29 FIRST-detected sources, we obtain an average radio luminosity L$\rm _{1.4GHz} \sim 5\times 10^{22}~W~Hz^{-1}$ (Fig.~\ref{power_size}). From the best-fit relation between 1.4 GHz radio continuum luminosity and the Balmer decrement corrected H$\alpha$ \citep{brown17}, we obtain a corresponding H$\alpha$ luminosity $\sim$ 2.5 $\times \rm 10^{42} \ erg\ s^{-1}$. Using the known relation between SFR and H$\alpha$ luminosity \citep{kennicutt09, brown17} assuming a Kroupa initial mass function (IMF) \citep{kroupa}, we obtain an expected star formation rate exceeding $\rm 5-10~M_{\odot}~yr^{-1}$, which is not observed in our galaxy sample. %From the best-fit relation between 1.4 GHz radio continuum luminosity and the Balmer decrement corrected H  (Brown et al. 2017), we obtain a corresponding H  luminosity 1:3 1041 erg s-1. Using the known relation between SFR and H  luminosity (Kennicutt et al. 2009; Brown et al. 2017) assuming a Kroupa intial mass function (IMF) (Kroupa & Weidner 2003), we obtain an expected SFR from this radio emission. 
  This is further confirmed by 
  Fig.~\ref{fig:quenched} (panel d) which shows that our objects lie above the low frequency radio luminosity - star formation rate relation from \cite{gurkan18}. This implies that the observed radio luminosity can not be explained by the detected very low amount of star formation and is consistent with radio emission from central radio AGN.
  
    \begin{figure}[h!!]
   \centering
   \includegraphics[width = 0.46\textwidth]{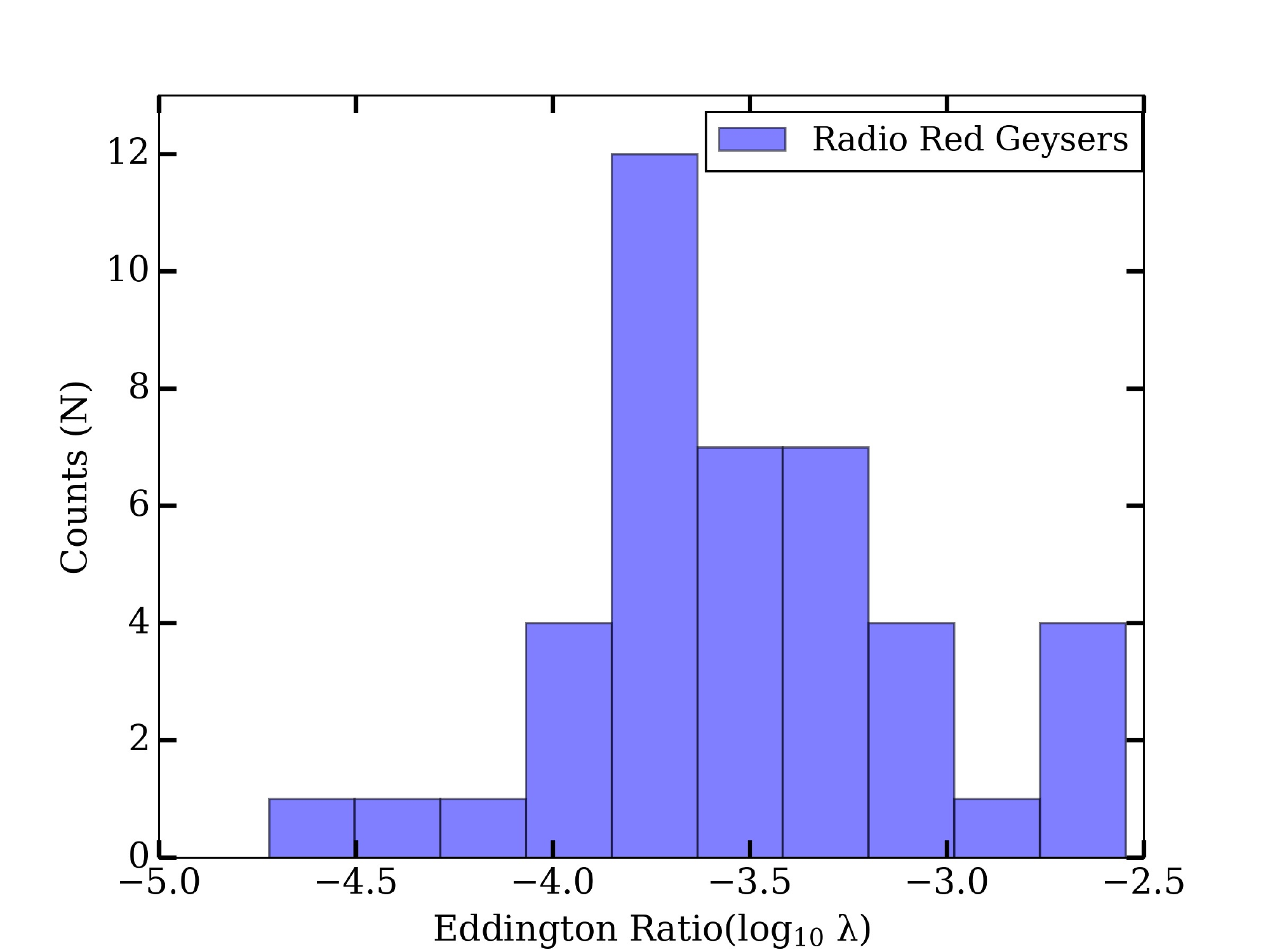}
   \caption{ The distribution of Eddington ratios for the 42 radio detected red geysers. The Eddington scaled accretion rate ($\rm \lambda$) values are $< 10^{-3}$ for the majority of the sources, implying that the red geysers are in general radiatively inefficient. }%
    \label{eddington}
    \end{figure}

    Another widely discussed source of the radio emission in
    radio-quiet sources is radiatively-driven accretion disc winds
    which result in synchrotron emitting shocks through the
    inter-stellar medium \citep{zakamska14, nims15, zakamska16}. 
    %(e.g. Jiang et al. 2010; Zakamska \& Greene 2014; Nims et al. 2015; Zakamska et al. 2016; Hwang et al. 2018). 
    However, these are generally associated with quasars with bolometric luminosity L$_{\rm AGN} \sim 10^{45}~\rm erg~s^{-1}$ where outflow velocities of $\sim \rm 1000~km~s^{-1}$ interact with the ISM and can produce radio luminosities similar to our sources (L$_{\rm 1.4 GHz} \sim 10^{22}-10^{23}~\rm W~Hz^{-1}$). However, the red geysers show typical gas velocities $\sim \rm 300~km~s^{-1}$ (Fig.~\ref{fig:geysers_eg}) with no signature of quasar-like broad emission lines. A rough  estimation of bolometric radiative luminosity from the [OIII] $\lambda$ 5007 \AA~emission line flux (within the central 2$''$ radius aperture), using the relation $\rm L_{rad} = 3500 L_{[OIII]}$ \citep{heckman04}, yields L$_{\rm rad} \sim 10^{43}~\rm erg~s^{-1}$. We calculate the classical eddington limit with $\rm L_{Edd} = 3.3 \times 10^{4} \ M_{BH}$, where $\rm M_{BH}$ has been calculated in \S\ref{ionized_1} (Eq. \ref{m_bh}). Inserting the jet mechanical energy E$_{\rm mech}$(Eq. \ref{e_mech}), radiative luminosity L$_{\rm rad}$ and eddington luminosity $\rm L_{Edd}$, we calculate the eddington radio $\rm \lambda = (E_{mech} + L_{rad})/ L_{Edd}$. Fig.~\ref{eddington} shows the distribution of Eddington ratios for the radio red geysers, which spans primarily between $\sim \rm 10^{-4} -  10^{-3}$. These fairly low eddington ratios implies that these are radiatively inefficient sources that cannot be formed due to radiative accretion disk winds.  
    %from One prediction

    Low-luminosity jets from  radio-quiet AGN are the most plausible explanation for the observed low power radio emission in our red geyser targets. Sufficiently deep and high resolution radio observations have been able to identify small-scale radio jets in several ``radio-quiet'' quasars and Seyfert galaxies \citep{gallimore06, baldi18, jarvis19, jarvis21, venturi21}. Our sample of radio-quiet red geysers have
    many properties in common with jetted compact radio galaxies %showing hot spots, jets, cores and compact features, 
    similar to those in \cite{kimball11, baldi18, jarvis19}. 
    
    \subsection{Radio red geysers in the context of other radio-quiet sources in the literature}
    
   At the frequencies of interest in this work, the radio continuum emission is dominated by non-thermal synchrotron emission. In the absence of a constant energy injection source, the replenishment of fresh electrons ceases and the radio spectrum is dominated by radiative loss. Since the energy loss rate is directly proportional to the frequency, the energy loss rate at lower frequencies ($\leq \rm 1.4 GHz$) is lower which enables the original injection index of the electrons to be retained for much longer. Thus, the lower frequency emission generally is more extended, characterizing emission from older plasma where injection took place longer ago. Additionally, LOFAR is more sensitive to extended emission than the FIRST and VLASS surveys. This is consistent with our findings (Fig.~\ref{compare}) which shows that the LOFAR measured sizes are roughly two-three times more spatially extended than the FIRST and VLASS images, sometimes revealing intriguing structures not visible in higher frequency images.

   We observe a range of radio morphologies from LOFAR observations in our red geyser targets. From Table~\ref{tab:summary}, we see that 16 out of 42 radio-detected sources exhibit resolved but compact morphology, with the spatial extent ranging between 3$-$7 kpc in 150 MHz frequency band. They can be represented by a two dimensional gaussian with no particular feature in their radio images. On the other hand, 16 other sources show extended features ($\ge$ 9 kpc), with contiguous one-sided morphology, bubbles and double-lobes (belonging to "extended", "irregular", "double" and "triple" morphology class). The remaining ten sources are unresolved in the typical resolution of LOFAR and FIRST. Thus, $\sim 38$\% of the radio-detected sources show spatially extended features extending to scales of more than ten kpc. This is consistent with the results of \cite{pierce20}, who have studied similar moderate luminosity radio AGN population, although with a greater radio luminosity range ($\rm 22.5 < L_{1.4GHz} < 25.$) than the red geysers. However, among those 38\%, the seven galaxies classified as ``extended'' also do not show any resolved radio jets although they show elongation in a specific direction, indicating some underlying radio structures remaining unresolved at the current spatial resolution.

   To quantitatively compare the radio morphologies of our sample to the traditional radio AGN population, we investigate the radio size versus radio luminosity plane for the red geysers compared to the literature compilation of radio selected AGN from \cite{an12, gallimore06, kukula98, mingo19, jarvis19} in Fig.~\ref{power_size}. In terms of linear size, most of the red geyser sources are similar to the compact steep spectrum sources (spanning $\sim \rm 1-25~kpc$ in 1.4 GHz radio image) with a few showing even more compact structures with no structures resolvable beyond the nuclear component, similar to the gigahertz peaked spectrum sources \citep[typically $<$ 1 kpc,][]{odea98}. However, unlike the red geysers, the CSS and GPS sources are powerful radio-loud AGN (L$_{\rm 1.4GHz} > 10^{25}\ \rm  W \ Hz^{-1}$). 
   Hence, our sources would be excluded from these samples due to a much lower radio-luminosity. The red geysers are more aligned with the ``radio-quiet quasars'' from \cite{jarvis19} and also from the LINER and Seyfert radio sources from the LeMMINGs survey \citep{baldi18b,baldi21} according to the radio luminosity- size diagram. A few are consistent with the lowest luminosity AGN in the sample of \cite{gallimore06}.

   As noted by \cite{jarvis19,jarvis21}, objects in the radio-quiet quasar (RQQ) category possess small-scale jets when observed in higher resolution (sub-arcsec scale) VLA and e-MERLIN images, but exhibit compact or slightly extended kpc-scale structures when observed in low spatial resolution ($> 3''$). They have similar radio morphology, spectral index and radio size as the red geysers. Considering the stark similarities of the radio properties of the red geyser sources with the RQQ sample, it seems plausible that our sources also possess small-scale jets which are blurred in the current resolution. %There have been several hypothesis to explain the lack of large scale jets in these compact radio-quiet populations. The most predominant one is  %relevant one  -  they are `young' sources that will eventually evolve to become traditional FRI or FRII radio galaxies \cite{odea98, hardcatle20}, or  
   %that they
These could resemble ‘frustrated’ jets, which are small and contained within the inner 1 kpc central region of the galaxy, occurring due to the surrounding dense environments that doesn't enable the jet 
to grow to a large size \citep{van84}.
   Two of our targets overlap with the Fanaroff-Riley class I \cite[FRI;][]{fanaroff74} galaxies in the luminosity–size plane 
   but the majority of our targets do not fit within the traditional FRI and FRII radio classifications. Due to the abundance of compact and featureless radio
morphology in our red geyser galaxies, these sources can also be classified as `FR0' galaxies \citep[e.g.][]{baldi15, capetti20a}.
However, higher spatial resolution data may reveal more complex
morphologies, jets, hot-spots on smaller scales \cite[see discussion in][]{hardcastle20}.

%   The jet interpretation is particularly strong due to the presence of compact, flat spectrum components (i.e., $\alpha \geq$  -0.5; likely to be hot spots; see e.g. Meisenheimer et al. 1989; Carilli et al. 1991) inside the more diffuse radio lobes. In a few cases, we see that the brightest nuclear radio component has a moderately flat spectral index (i.e.,$\alpha \geq$  -0.5), which may indicate a contribution from radio emission associated directly with an AGN 'core' / accretion disc (Padovani 2016).

\subsection{Interaction between radio and ionized gas}
 
   We have previously identified kpc-scale outflows in ionized gas in the red geyser galaxies \citep{cheung16, roy20}, which are marked by the bi-symmetric extended pattern in the H$\alpha$ EW map (Fig~\ref{fig:geysers_eg}). Our study of red geysers complements several other studies that have aimed to characterize the drivers of ionised outflows by investigating the radio properties of the central radio AGN. For example, there
are many spatially-resolved studies of multiphase gas outflows driven by local galaxies hosting low power AGN \citep{mingozzi19, wylezalek20,  venturi21, dutra21}.
%( Mingozzi et al. 2019;  Wylezalek et al. 2020; Davies et al. 2020a; Venturi et al. 2020). 
Specifically, \cite{capetti19, capetti20a} have studied FR0 galaxies showing similar compact radio structures detected with LOFAR observations and their effect on the galaxy environments, while \cite{webster21} has discovered galaxy scale jets and their interaction with the interstellar medium of the host galaxy. There have been numerous studies at high redshift as well, but they primarily focus on powerful radio galaxies and luminous AGN \citep{nesvadba17, circosta18, perna15, zakamska16}
%(Nesvadba et al. 2017, Circosta et al. 2018, Perna et al. 2015, Zakamska et al. 2016b) 
where the impact of quasars are dominant \citep{hopkins07}. 
 The faint radio emission from radio-quiet AGN are difficult to detect, although a growing number of studies are being done in recent times to study the interaction between the radio emission from the AGN and the ionized gas outflows \citep{alyazeedi21}. However, high signal-to-noise and high spatial resolution
radio data \citep[e.g.,][]{jarvis19, venturi18}
 are required to resolve the small-scale radio jets and morphologies to establish further connection with the ionized outflows.

%%%
   
   Fig.~\ref{halpha_size} aims to establish the connection between ionized gas and radio emission in the red geysers via integrated H$\alpha$  luminosity with radio sizes. We show that our results are consistent with \cite{jarvis21} studying similar radio-quiet ($\rm L_{150MHz} < 10^{24}\ W\ Hz^{-1}$) compact (generally $< \rm 10 \ kpc$) AGN sources from the ``Quasar-feedback survey'', which states that ionized gas tracers are correlated with the central radio AGN. 
   Additionally, radio detected red geysers not classified as ``compact'', possessing large radio sizes, are shown to have relatively lower sSFR ($\rm log_{10}\ sSFR \sim -13.0\ yr^{-1}$) than the compact ones (Fig.~\ref{ssfr_l150}). Assuming these compact objects have similar star formation history as the extended, lobed \& irregular ones,
large radio sources having lower SFR for a given stellar mass might be implying more efficient quenching due to the presence of jets than the compact ones. 
The other possibility is that the extended radio sources are generally found in more massive and evolved sources with higher stellar mass, perhaps even massive central galaxies in large halos. Extended or ``lobed'' radio sources may be the evolved form of compact ``FR0'' galaxies in such environments. This idea is consistent with the theory that compact FR0 radio galaxies are younger and will eventually evolve to form the traditional FRI or FRII galaxies \citep{odea98}. Thus the range of radio morphologies observed in the red geysers could represent the various stages of transition of FR0 class of objects to the traditional FRI/ FRII sources. 
% However, the high fraction of compact radio morphologies indicate that not all of them will evolve into extended radio sources. The two primary factors responsible for that are intrinsic differences in the central engines and the wide range of environments in the host galaxies. Small scale slow jets are subject to instabilities and they disrupt quickly leading to their inability to extend beyond the host galaxy.   

   Fig.~\ref{fig:ionized} and \ref{fig:ionized2} compare the spatially resolved equivalent width and kinematics of ionized gas, traced by H$\alpha$, compared to the distribution of radio emission for five example red geysers belonging to different radio morphological class. %We overlay our radio images from LoTSS survey on top of the kinematics maps from the IFS data to understand the interaction and connection between radio features and ionized gas. For the galaxies classified under ``compact'' and ``extended'' category, we can not infer about the spatial correlation between radio-jets and ionized wind due to the absence of visible radio jets (see MaNGAID: 1-245451 in Fig.~\ref{fig:ionized} for an example). However, 
   We find that for the galaxies in the ``irregular'' and ``double'' class, elongation along a specific direction or presence of one-sided bubbles in the radio emission (MaNGAID: 1-188530 and 1-23958, Fig.~\ref{fig:ionized}) roughly aligns with the ionized gas features, marked by high gas dispersion. Such observations indicate that the large scale ionized outflows, stretching to $> \rm 10\ kpc$, are possibly driven by small scale radio jets, unresolved at the current LOFAR resolution. Often in these cases, the velocity dispersion map is clumpy and shows high values in distinct parts of the galaxy, and also in some cases, perpendicular to the bi-symmetric H$\alpha$ feature. 
    This is similar to the observation by \cite{venturi21} who found that increased line widths were perpendicular to small-scale radio jets. They interpreted this to be due to the low-power jet strongly interacting with the ISM in the galaxy, releasing energy and giving rise to highly turbulent motions in the perpendicular direction. Similar characteristics have been seen in a few other local seyfert galaxies as well \citep[see ][]{riffel14, riffel15, lena15, freitas18}.

 However, among the three rare cases (classified as ``triple'') where we detect clear evidence of large scale radio jets and lobes, particularly notable are the galaxies with MaNGAID: 1-378770 and 1-595166 where we find that the radio jets, shown by LOFAR radio contours, and ionized wind, traced by the H$\alpha$ emission map, lie  perpendicular to each other (Fig.~\ref{fig:ionized2}).

   There can be several possible explanations for the perpendicular incidence of the radio jet and ionized broad angled wind. 
   \begin{itemize}
       \item Changing orientation of the magnetic field, as proposed by 
       \cite{mehdipour19}. 
       \item The formation of an expanding cocoon structure described in simulations of jet-driven feedback \citep{begelman84}.  
       \item A precessing accretion disk due to a misalignment between the orientation of the disk and the spin of the black hole \citep{riffel19}.
       \item Radiation driven midplane wind \citep{proga04}.
   \end{itemize}

 In a sample of 16 radio loud Seyfert-1 AGN galaxies, \cite{mehdipour19} showed an inverse correlation between the column density of the ionized wind and the radio loudness parameter (R) of the jet observed mis-aligned with the wind. They argued that this indicates a wind-jet bimodality in radio loud AGNs with the AGN alternating between powering a radio jet and an un-collimated broad wind. They proposed that the magnetic field is the primary driving mechanism for the observed accretion-disk wind and the change in the magnetic field configuration from toroidal to poloidal cause this switch. This bi-modality can explain the low incidence of radio jets and high prevalence of ionized wind in the red geyser sample along with a 90$^{\circ}$ mis-alignment between the jet and the wind. However, if the magnetic driving mechanism can work equally efficiently on low power radio-quiet AGNs is still open to questioning.

  \cite{wagner11}, \cite{wagner12}, \cite{mukherjee16} and \cite{mukherjee18} have studied the interaction between radio-jet and multi phase ISM using detailed 3D hydro-dynamical simulations. They find that when a radio jet propagates through an inhomogenous ISM, they not only impact the ISM along the radio-jet axis but also create a spherical bubble which drives gas clouds outwards in all directions. This leads to outflowing gas traveling at a modest speed mostly in the path of least resistance. Although this mechanism can not lead to extreme velocity outflows escaping the galaxy altogether, it can impart enough energy and turbulence to heat the gas and sufficiently inhibit star formation. Since we observe similar modest outflowing gas velocities ($\sim 300~\rm km~s^{-1}$) accompanied by turbulent high velocity dispersions (exceeding $\sim 220~\rm km~s^{-1}$) in specific regions of the galaxy in red geysers with very little star formation activity, this phenomenon might be the primary mechanism to explain the  perpendicularity between radio jets and winds observed in the two red geysers. 
   
   A third possibility is due to the precession in the accretion disk, as proposed by \cite{riffel19}. They suggested that if there is a misalignment between the orientation of the accretion disk and the spin of the black hole, it can create a torque leading to a constant precession. This mechanism, in turn, can lead to a small scale jet, contained within the central part near the nucleus to be mis-aligned 
   with the large scale ionized wind observed in MaNGA. \cite{riffel19} has successfully implemented this model to the prototypical red geyser to explain the misalignment observed between the direction of the ionized wind in small ($1-2$kpc) and large ($>10$ kpc) scale spatially resolved H$\alpha$ map. A similar procedure can possibly explain the large scale radio jet and ionized wind mis-alignment, although further work is needed to confirm whether the small-scale precession near the black hole can be implemented in a larger spatial scale $>20 $ kpc. 
   
   The final proposed mechanism exploring the radiation driven mid-plane disk wind has been proposed by \cite{proga04}. However, as discussed in \S\ref{discussion_1}, the red geysers do not seem to harbor radiatively-driven accretion disk winds owing to low eddington ratios between $\rm 10^{-4}-10^{-3}$ and absence of extreme outflow velocities of $\sim \rm 1000\ km\ s^{-1}$.

   Finally, while there is a body of evidence that supports the interpretation of red geyser kinematics as the result of outflowing winds, this interpretation may be wrong in some cases.  It is tempting to associate the perpendicular orientation of the major kinematic axis with a diffuse accretion disk, giving rise to extended, bipolar radio jets.  While difficult to rule out, this explanation is unlikely because it implies that a black hole accretion disk on sub-parcsec scales would remain aligned to a galaxy-scale gaseous disk on kpc scales.  In simulations and observations, such an alignment is extremely rare.
   
%   In summary, both winds and low-power jets will take the path of least resistance while moving through the clumpy ISM medium, that can result in a misalignment although the exact mechanism is still not clear. However, our observations, both integrated and resolved, have provided strong evidence for an impact of the radio jets on the surrounding interstellar medium. Similar evidences have been observed in \cite{jarvis19, venturi21} who have combined the power of spatially resolved studies in both radio properties and multi phase gas kinematics. A better understanding of the red geysers in particular would require radio and optical observations with higher ($< 1''$) spatial resolution to resolve the smaller scale radio jets and corresponding turbulence and shock signatures in ionized gas. 

   \section{Conclusion} \label{conclusion}
   
   We have studied the 150 MHz, 1.4 GHz and 3 GHz radio images from LoTSS, FIRST and VLASS surveys ($\rm 6'', 5'' and \ 2.5''$ resolution respectively) along with integral field spectroscopic observations of a sample of red geyser galaxies. Red geysers are low redshift (z$<$0.1) passive early-type galaxies that host ionized gas outflows on scales of $\sim $10 kpc \citep{cheung16, roy20}. The parent sample of 140 red geysers are selected from SDSS IV-MaNGA MPL-9 data. 42 out of the total 140 red geysers are detected in at least one of the three radio surveys while 21 sources are detected in all three surveys. We present the radio characteristics, morphology, size of the radio-detected red geysers and explore the connection of the radio emission with ionized gas. Our main conclusions are:
   
   \begin{itemize}
       \item Only 3-4 of the 42 radio detected red geysers are radio-loud according to two traditional criteria \citep{xu99, kellermann89}. Red geysers are largely low luminosity sources and are classified as ``radio-quiet'' objects.
       \item Although these are radio-quiet, the source of the detected radio emission is the central radio AGN and not star formation. We use a series of four diagnostic diagrams (see Fig.~\ref{fig:quenched}) to show the absence of sufficient star formation to explain the observed radio luminosity. The association with the central radio AGN is confirmed by Fig.~\ref{jet_smbh} which shows a moderately tight correlation of the jet mechanical energy derived from radio luminosity with the SMBH mass.
       
       \item 16 out of 42 sources show extended radio structures with a diverse range of morphologies, with radio sizes spanning a large range $\rm 9\ kpc < size < 200\ kpc$. Two sources are classified as "FRI" sources. The remaining 26 sources are either unresolved or exhibit resolved but compact structure with size $\rm  < 9\ kpc$. The compact sources have no particular feature in their radio image.
       \item Based on their radio luminosity-size relationship, spectral index and the observed radio morphology, these galaxies are consistent with ``radio-quiet'' quasars \citep{jarvis19}, low power compact radio galaxies called the ``FR0'' sources \citep{baldi15, capetti20a} and the radio emitting LINERs and Seyfert class \citep{gallimore06, baldi21}. Higher resolution ($<\rm 1''$) radio images are required to detect small scale ``frustrated'' radio jets within these compact sources, if there are any. 
       \item We show that there are indications of interaction between the radio structures and the ionized gas (traced by H$\alpha$). Specifically, H$\alpha$ luminous sources tend to have more extended radio emission, in general (Fig.~\ref{halpha_size}). 
       \item We find evidence that compact radio red geysers show a higher specific star formation rate on average than those possessing large extended radio structures (Fig.~\ref{jet_smbh}). This could mean larger radio sources with visible lobes and jets are more efficient in quenching than the compact ones and thus having less SFR. This could also mean that the extended radio sources are generally found in more massive and evolved sources with higher stellar mass, causing them to have the lowest sSFR. The later possibility could imply that the compact `FR0' galaxies would eventually evolve to transition towards jetted FRI/FRII sources. Thus, the range of radio morphologies observed in the red geysers represent the various stages of this transition.  
       \item From spatially resolved maps, the ionized gas and the radio
structures are mostly co-spatial with distinct kinematic
components. However, in two out of the three objects where we detected large scale radio lobes, the ionized wind and the radio lobes are perpendicular to each other. These can arise from the jet-ISM interaction via different mechanisms (see \S\ref{discussion}). 
       
   \end{itemize}
   
   In this work we provide evidence that the compact radio structures are a common characteristic feature of red geyser galaxies. 
   In order to test the presence of small scale radio jets within the compact morphology and to study the detailed radio jet-ISM interaction, further higher-resolution (sub-arcsecond) radio imaging from VLA and upcoming more sensitive and powerful radio and optical telescopes like Next Generation Very Large Array (ngVLA), Square Kilometer Array (SKA) and Vera C. Rubin Observatory will be essential.

%%%%%%%%%%%%%%%%%%%%%%%%%%%%%%%%%%%%%%%%%

\section*{Acknowledgements}
This research was supported by the National Science Foundation under Award No. 1816388. The authors thank the anonymous referee for helpful comments that significantly improved the manuscript. 

LOFAR, the Low Frequency Array designed and constructed by ASTRON, has
facilities in several countries, which are owned by various parties
(each with their own funding sources), and are collectively operated
by the International LOFAR Telescope (ILT) foundation under a joint
scientific policy. The ILT resources have benefited from the
following recent major funding sources: CNRS-INSU, Observatoire de
Paris and Universit\'e d'Orl\'eans, France; BMBF, MIWF-NRW, MPG, Germany;
Science Foundation Ireland (SFI), Department of Business, Enterprise
and Innovation (DBEI), Ireland; NWO, The Netherlands; the Science and
Technology Facilities Council, UK; Ministry of Science and Higher
Education, Poland.

Part of this work was carried out on the Dutch national
e-infrastructure with the support of the SURF Cooperative through
grant e-infra 160022 \& 160152. The LOFAR software and dedicated
reduction packages on \url{https://github.com/apmechev/GRID_LRT} were
deployed on the e-infrastructure by the LOFAR e-infragroup, consisting
of J.\ B.\ R.\ Oonk (ASTRON \& Leiden Observatory), A.\ P.\ Mechev (Leiden
Observatory) and T. Shimwell (ASTRON) with support from N.\ Danezi
(SURFsara) and C.\ Schrijvers (SURFsara). This research has made use of the University
of Hertfordshire high-performance computing facility
(\url{https://uhhpc.herts.ac.uk/}) and the LOFAR-UK compute facility,
located at the University of Hertfordshire and supported by STFC
[ST/P000096/1]. The J\"ulich LOFAR Long Term Archive and the German
LOFAR network are both coordinated and operated by the J\"ulich
Supercomputing Centre (JSC), and computing resources on the
supercomputer JUWELS at JSC were provided by the Gauss Centre for
supercomputing e.V. (grant CHTB00) through the John von Neumann
Institute for Computing (NIC).

E.M. acknowledges financial support from the Czech Science Foundation project No.19-05599Y. This work was supported by the EU-ARC.CZ Large Research Infrastructure grant project LM2018106 of the Ministry of Education, Youth and Sports of the Czech Republic.

RR thanks Conselho Nacional de Desenvolvimento Cient\'{i}fico e
Tecnol\'ogico  ( CNPq, Proj. 311223/2020-6,  304927/2017-1 and
400352/2016-8), Funda\c{c}\~ao de amparo 'a pesquisa do Rio Grande do
Sul (FAPERGS, Proj. 16/2551-0000251-7 and 19/1750-2),
Coordena\c{c}\~ao de Aperfei\c{c}oamento de Pessoal de N\'{i}vel
Superior (CAPES, Proj. 0001).

RAR acknowledges support from Conselho Nacional de Desenvolvimento Cient\'ifico e Tecnol\'ogico (CNPq) and Funda\c c\~ao de Amparo \`a Pesquisa do Estado do Rio Grande do Sul (FAPERGS). 

Funding for the Sloan Digital Sky Survey IV has been provided by the Alfred P. Sloan Foundation, the U.S. Department of Energy Office of Science, and the Participating Institutions. SDSS-IV acknowledges
support and resources from the Center for High-Performance Computing at
the University of Utah. The SDSS web site is \href{http://www.sdss.org}{www.sdss.org}.

SDSS-IV is managed by the Astrophysical Research Consortium for the 
Participating Institutions of the SDSS Collaboration including the 
Brazilian Participation Group, the Carnegie Institution for Science, 
Carnegie Mellon University, the Chilean Participation Group, the French Participation Group, Harvard-Smithsonian Center for Astrophysics, 
Instituto de Astrof\'isica de Canarias, The Johns Hopkins University, 
Kavli Institute for the Physics and Mathematics of the Universe (IPMU) / 
University of Tokyo, the Korean Participation Group, Lawrence Berkeley National Laboratory, 
Leibniz Institut f\"ur Astrophysik Potsdam (AIP),  
Max-Planck-Institut f\"ur Astronomie (MPIA Heidelberg), 
Max-Planck-Institut f\"ur Astrophysik (MPA Garching), 
Max-Planck-Institut f\"ur Extraterrestrische Physik (MPE), 
National Astronomical Observatories of China, New Mexico State University, 
New York University, University of Notre Dame, 
Observat\'ario Nacional / MCTI, The Ohio State University, 
Pennsylvania State University, Shanghai Astronomical Observatory, 
United Kingdom Participation Group,
Universidad Nacional Aut\'onoma de M\'exico, University of Arizona, 
University of Colorado Boulder, University of Oxford, University of Portsmouth, 
University of Utah, University of Virginia, University of Washington, University of Wisconsin, 
Vanderbilt University, and Yale University.\\

%% ------------------------------------------------------------------
%% REFERENCES 
%% 
%% ------------------------------------------------------------------
%\newpage

\end{document}